\documentclass[lettersize, journal]{IEEEtran}
\usepackage{amsmath,amsfonts}
\usepackage{algorithmic}
\usepackage{algorithm}
\usepackage{caption}
\usepackage{subcaption}
\usepackage{array}
\usepackage{array, multirow, graphicx}
\usepackage{multicol}
\usepackage{textcomp}
\usepackage{stfloats}
\usepackage{url}
\usepackage{verbatim}
\usepackage{graphicx}
\usepackage{cite}
\usepackage{xcolor}
\usepackage{hyperref}
\usepackage{multirow}
\begin{document}

\title{GPS-IDS: An Anomaly-based GPS Spoofing Attack Detection Framework for Autonomous Vehicles}

%\author{IEEE Publication Technology,~\IEEEmembership{Senior Member,~IEEE,}

\author{Murad Mehrab Abrar, Amal Yousseef, Raian Islam, Shalaka Satam, \\ Banafsheh Saber Latibari, 
Salim Hariri, Sicong Shao, Soheil Salehi, and Pratik Satam

\thanks{This work is partly supported by National Science Foundation (NSF) research projects NSF-1624668, (NSF) OIA-2218046, (NSF) FAIN-1921485 (Scholarship-for-Service), NSF-2213634, Department of Energy/National Nuclear Security Administration under Award Number DE-NA0003946, AGILITY project 4263090 sponsored by KIAT (South Korea), and the University of Arizona's Research, Innovation \& Impact (RII) award for the ``Future Factory".}
}
% <-this % stops a space
%\thanks{Manuscript received April 19, 2021; revised August 16, 2021.}}

\markboth{IEEE TRANSACTIONS ON DEPENDABLE AND SECURE COMPUTING}%~Vol.~14, No.~8, August~2021}
{Mehrab Abrar \MakeLowercase{\textit{et al.}}: GPS-IDS: An Anomaly-based GPS Spoofing Attack Detection Framework for Autonomous Vehicles}


\maketitle

\begin{abstract}

Autonomous Vehicles (AVs) heavily rely on sensors and communication networks like Global Positioning System (GPS) to navigate autonomously. Prior research has indicated that networks like GPS are vulnerable to cyber-attacks such as spoofing and jamming, thus posing serious risks like navigation errors and system failures. These threats are expected to intensify with the widespread deployment of AVs, making it crucial to detect and mitigate such attacks. This paper proposes GPS Intrusion Detection System, or GPS-IDS, an Anomaly-based intrusion detection framework to detect GPS spoofing attacks on AVs. The framework uses a novel physics-based vehicle behavior model where a GPS navigation model is integrated into the conventional dynamic bicycle model for accurate AV behavior representation. Temporal features derived from this behavior model are analyzed using machine learning to detect normal and abnormal navigation behaviors. The performance of the GPS-IDS framework is evaluated on the AV-GPS-Dataset--- a GPS security dataset for AVs comprising real-world data collected using an AV testbed, and simulated data representing urban traffic environments. %The dataset has been publicly released for the global research community.
To the best of our knowledge, this dataset is the first of its kind and has been publicly released for the global research community to address such security challenges.

\end{abstract}

\begin{IEEEkeywords}
Autonomous Vehicles, Anomaly Detection, CARLA, GPS Spoofing, Intrusion Detection System, Physics-based Behavior Modeling, Robotics, Secure Navigation. 
\end{IEEEkeywords}

\section{Introduction}\label{sec1}
\IEEEPARstart{V}{ehicles} are getting increasingly autonomous and becoming ever more reliant on onboard sensors and communication networks, aiming to make transportation faster, safer, and environmentally sustainable by reducing human intervention in driving tasks \cite{SAE International, shalaka dissertation}. Researchers have shown human error to be the cause of over 90\% accidents, resulting in 40 thousand deaths and over 2 million injuries annually in the United States alone \cite{Critical reasons for crashes, Early Estimates of Motor Vehicle Traffic Fatalities} --- a statistic that will be reduced with the transition to AVs \cite{Traffic accidents with autonomous vehicles}. AVs rely on \textit{Automotive Sensing}, a collection of diverse sensors that enable environmental perception and safe navigation without requiring constant human input \cite{Autonomous vehicle perception}. Automotive sensing is broadly categorized into three types: \textit{Self-sensing}, \textit{Surrounding-sensing}, and \textit{Localization} \cite{Autonomous vehicle perception}. \textit{Self-sensing} refers to how an autonomous vehicle gathers and interprets information about its own state, including its position, velocity, and acceleration. \textit{Surrounding-sensing} is the ability of a vehicle to perceive its environment, like recognizing traffic signs, understanding weather conditions, or measuring the state of other vehicles around it. \textit{Localization} %determines the local and global positions of the vehicle and helps it to navigate safely to the desired destination. 
integrates internal self-sensing and surrounding sensing data with external information from satellite-based navigation systems to determine the local and global positions of the vehicle. For self-sensing and surrounding sensing, AVs use an array of sensors like Inertial Measurement Units (IMUs), gyroscopes, odometers, Controller Area Network (CAN) bus, cameras, Light Detection and Ranging (LiDARs), etc. \cite{Artificial intelligence applications in the development of autonomous vehicles}. For localization and navigation, they rely on satellite-based navigation systems like Global Navigation Satellite System (GNSS) \cite{Autonomous driving}. 

GNSSs like GPS (United States), GLONASS (Russian) \cite{glonass}, BeiDou (China) \cite{beidou}, and Galileo (European Union) \cite{galileo}, provide geolocation and time information to a receiver anywhere on or near the Earth. Being the pioneering system, GPS utilizes a 24-satellite constellation to provide its users with highly precise and accurate location and time information for navigation \cite{on the requirement of successful gps spoofing}. GPS offers two distinct variants: a secure military-grade GPS that is exclusively accessible to the United States military branches and a civilian GPS that is available for public use \cite{intelligent detection of gps attack}. Most autonomous systems, including autonomous vehicles, robots, and drones, rely on the latter variant for navigational purposes. This heavy reliance on civilian GPS raises significant concerns due to the absence of encryption or authentication mechanisms, unlike the military-grade GPS \cite{assessing the spoofing threat}. Researchers have demonstrated that civilian GPS is vulnerable to jamming and spoofing attacks, and commercially available off-the-shelf GPS receivers lack the capability to detect and counteract such attacks \cite{on the requirement of successful gps spoofing, drift with devil, assessing the spoofing threat, inside GNSS}. Various GPS spoofing techniques, including Lift-off-delay \cite{lift off delay}, Lift-off-aligned \cite{lift off aligned}, Meaconing or Replay \cite{lift off aligned}, Jamming and Spoofing \cite{jamming and spoofing}, and Trajectory Spoofing \cite{trajectory spoofing}, pose serious threats to the integrity of GPS. Such spoofing attacks can result in navigational errors, potential vehicle hijacking, or fatal collisions. %Therefore, it is crucial that modern autonomous vehicles integrate a GPS attack detection system that is more effective, accurate, built upon real-life navigational scenarios, and provides a comprehensive understanding of GPS attacks from a holistic perspective.

Motivated by these challenges, this paper presents GPS-IDS, an Anomaly-based Intrusion Detection System that uses a novel physics-based vehicle behavior model (a modification of the conventional dynamic bicycle model) and machine learning to detect GPS spoofing attacks on AVs. The bicycle model is a simplified representation of a vehicle's dynamics, typically used in the context of autonomous vehicle control systems. It approximates the vehicle as a two-wheel system, capturing the essential dynamics without the complexity of a full multi-wheel model. %This simplification helps in understanding and developing control strategies for vehicle navigation and stability. 
By extracting temporal features from this model, GPS-IDS captures the normal behavior of AVs. Machine learning models are then utilized to differentiate between normal and abnormal behaviors. The detection of GPS spoofing attacks is achieved by identifying deviations from the normal behavior pattern, indicating potential intrusions.

The main contributions of this paper
are as follows:

\begin{itemize}

   %\item This paper presents a modified dynamic bicycle model called the Autonomous Vehicle Behavior Model that integrates an autonomous GPS navigation model into a dynamic bicycle model, thus allowing vehicles' lateral dynamics and state space representations to capture normal behavior.

   \item The paper presents the \textit{Autonomous Vehicle Behavior Model}--- a hybrid vehicle model that integrates a modified physics-based dynamic bicycle model with GPS navigation. This modification allows for an accurate representation of the normal navigation behavior of AVs. Temporal features extracted from the physics-based model are utilized in the data-centric analysis to detect abnormal behaviors using machine learning.

   \item The paper introduces the \textit{AV-GPS-Dataset}, including: 

   \begin{itemize}
        \item \textit{Real-world Data:} Collected from field experiments with real GPS spoofing attacks on an AV testbed, capturing 44 features of GPS-guided navigation.
        \item \textit{Simulation-generated Data:} Generated using the CARLA (Car Learning to Act) \cite{CARLA} simulator, representing various urban scenarios with diverse traffic conditions. This part of the dataset helps to understand GPS spoofing effects in urban environments.
    \end{itemize}

    \item In contrast to the existing related datasets collected solely from simulated environments \cite{A survey of anomaly detection for connected vehicle, Whelan_Novelty-based intrusion detection, UAV Attack Dataset}, the AV-GPS-Dataset features both real-world and simulated data, providing normal and GPS spoofing attack scenarios.

    \item The GPS-IDS framework is evaluated using the AV-GPS-Dataset. The framework achieves:
    \begin{itemize}
        \item An F1 score of up to 94.4\%, with a 56.5\% improvement in detection time compared to the existing EKF-based GPS/INS detector in real-world data.
        \item An F1 score of up to 97.1\% in simulation-generated data with urban traffic scenarios, demonstrating high effectiveness in diverse environments.
    \end{itemize}
\end{itemize}
   
   %consists of two parts: a real-world dataset, and a simulation-generated dataset using CARLA (Car Learning to Act) \cite{CARLA}. In contrast to the existing related datasets collected from only simulated environments \cite{A survey of anomaly detection for connected vehicle, Whelan_Novelty-based intrusion detection, UAV Attack Dataset}, GPS-IDS features both real-data collected from practical field experiments with real-world GPS spoofing attacks performed on an autonomous vehicle testbed, and a simulation generated data representing urban environment and scenarios with diverse traffic conditions using. Both dataset includes normal and GPS spoofing attack data.
.

The rest of the paper is organized as follows: Section \ref{related work} discusses the background and related work; Section \ref{gps ids} introduces the GPS-IDS framework and explains the components; Section \ref{attacker model} explains the attacker model and defines the problem statement, Section \ref{experiments} presents the experimental evaluation of the GPS-IDS framework along with the dataset details, and Section \ref{conclusion} concludes the paper and outlines future directions.

\section{Background and Related Work}\label{related work}

This section outlines the relevant technical background on vehicle modeling techniques, various types of Intrusion Detection Systems, existing works on cyber-attacks and GNSS security for vehicles, and our motivations to address the current limitations.

\subsection{Vehicle Models}
\subsubsection{Physics-based Model}
A physics-based vehicle model is a mathematical representation that simulates the vehicle's physical behavior using factors like kinematics, dynamics, motion, forces, and environmental interactions. In \cite{Adaptive EKF-based vehicle state estimation}, the authors adopted a physics-based dynamic vehicle model incorporating an EKF estimator to determine the vehicle’s longitudinal and lateral velocities and yaw rate. In \cite{Kinematic and dynamic vehicle models for autonomous driving control design}, Kong et al. explored two physics-based vehicle models, namely kinematic and dynamic bicycle models, for model-based controller design in autonomous driving and presented a comparison using experimental data. Several research contributions have employed software tools such as Modelica, Robot Operating System (ROS), Gazebo, and Simulink to simulate physics-based vehicle models or their components. %Notable applications include racing car modeling 
\cite{racing car modelica}, 
%vehicle drivability modeling 
\cite{modeling vehicle drivability}, %heterogeneous physical system modeling, and simulating AV testbeds 
\cite{cat vehicle}.

\subsubsection{Data-centric Model}

A data-centric vehicle model uses data-driven techniques and machine learning algorithms to represent the vehicle system. The model analyzes a large volume of data from sensors and sources to determine the vehicle's actions. In \cite{Data-driven vehicle modeling} the authors proposed a data-driven modeling approach based on Deep Neural Networks (DNNs) to compute and predict the dynamic characteristics of a vehicle. In \cite{data driven modeling and control}, a data-driven system identification and control method is proposed for autonomous vehicles, where several vehicle dynamic variables are measured and corresponding data is collected to model a physical vehicle. In \cite{Longitudinal vehicle dynamics}, the authors conducted a comparative analysis between physical and data-driven vehicle models under real-world driving conditions.

\subsection{Intrusion Detection System (IDS)}

Intrusion Detection Systems (IDS) are software designed to detect cyber-attacks and can be classified into four types \cite{survey of intrusion detection in iot}: 1) Signature-based IDS, 2) Anomaly-based IDS, 3) Specification-based IDS, and 4) Hybrid IDS. Signature-based IDSs rely on a pre-learned attack signature database, where each observation matched with an entry in the attack database is considered as an attack. Signature-based IDSs heavily rely on an up-to-date attack signature database and are unable to detect new or modified attacks \cite{WIDS}. An Anomaly-based IDS relies on modeling techniques to capture the system's normal behavior, wherein any observed behavior outside the model-defined norm is classified as malicious. This reliance on normal behavior models allows Anomaly-based IDSs to detect new or modified attacks at the cost of increased modeling complexity \cite{ABA for IoT sensors}. Specification-based IDSs use a set of rules and policies that define the expected behavior of different system components such as sensor nodes or motor commands. Hybrid IDSs are a combination of Signature-based, Anomaly-based, and Specification-based IDSs. The Anomaly-based IDS approach presented in this work detects GPS spoofing attacks on AVs by utilizing an extensive physics-based AV behavior model and real-life vehicular data, aiming to use a more accurate operational baseline and overcome the high error rates present in Anomaly-based IDSs.

\subsection{Cyber-attacks on Vehicles}

The increasing deployment of AVs increases the cyber-attack surfaces within the vehicle system that can be potentially exploited. In \cite{hoppe}, Hoppe et al. exploited vulnerabilities present in the CAN-bus of a vehicle to attack the windows, lights, and airbags. Similarly, Miller et al. \cite{miller} successfully attacked a Jeep Cherokee 2014 by reprogramming a gateway chip in the head unit of the vehicle. The attack enabled the vehicle to send arbitrary CAN messages, allowing access to different critical subsystems like braking and steering \cite{lessons learned}. Similar attacks have been demonstrated against Toyota Prius 2010 and Ford Escape 2010 \cite{remote exploitation}, by targeting different vehicular Electronic Control Units (ECU) and head units \cite{checkoway}. Similar to the CAN bus attacks, researchers have successfully targeted navigation systems in AVs like the GPS receiver  \cite{Psiaki and Humphreys}, leading to hijacking attacks.

\subsection{GNSS Security}\label{gnss security}

In this section, we highlight the research aimed at detecting GNSS attacks. GNSS security literature can be classified into two broad categories: (1) GPS signal characteristics-based approach \cite{Psiaki and Humphreys}, \cite{Liang Chen}, \cite{He}, \cite{anomaly detection cav}, \cite{Milaat and Liu}, \cite{Oligeri}, %\cite{Real-time GPS spoofing detection via correlation of encrypted signals, Kern and Humphreys},
and (2) Machine learning-based approach \cite{bib31}, \cite{bib37} \cite{bib32}, \cite{Whelan_Novelty-based intrusion detection}, \cite{bib35}, \cite{bib36}. The GPS signal characteristics-based approaches rely on signal processing techniques to detect attacks. For instance, in \cite{Psiaki and Humphreys}, Psiaki and Humphreys proposed a GPS spoofing attack detection scheme based on the direction-of-arrival (DOA) induction. He et al. \cite{He} proposed the use of GPS signal distortion to detect GPS spoofing attacks. For connected and autonomous vehicles, an anomaly detection model is proposed by Yang et al. \cite{anomaly detection cav} to detect GPS spoofing on the localization system using the ``Learning From Demonstration" technique. Milaat and Liu \cite{Milaat and Liu} proposed a decentralized technique where vehicles exchange GPS code pseudo-range values with neighbors using short-range communications to detect high correlations during spoofed GPS signal arrival.  On the other hand, machine learning-based approaches rely on machine learning and data analytics techniques to detect attacks. Researchers have used one-class classifiers  \cite{Whelan_Novelty-based intrusion detection}, Artificial Neural Networks \cite{bib31}, Long Short Term Memory (LSTM) networks \cite{bib32}, \cite{bib35} to detect GPS attacks.

The majority of these works adopt a segregated approach focused on sensor node-specific detection techniques and either use a small (and restrictive) GPS dataset or an attack dataset collected solely from simulated environments using tools like Matlab and Gazebo, lacking the representation of the real world. This paper overcomes these limitations by: \\
(1) Introducing an approach that is evaluated on both real-world data including practical GPS spoofing scenarios, as well as simulated data in diverse urban environment conditions, and (2) Employing a hybrid vehicle behavior model, a combination of physics-based and data-centric models, to accurately represent the behavior of AVs. The physics-based component is a modified dynamic bicycle model that includes GPS navigation, and the data-centric component extracts temporal features from this model to model the vehicle’s normal navigation behavior and identify the deviations using machine learning — thereby detecting GPS attacks. 

Unlike the traditional GNSS security approaches, our approach benefits from diverse data and the combination of both physics-based and data-centric models, giving the IDS a comprehensive view and allowing faster detection of anomalies like GPS attacks.

\begin{figure}[tp]
\centerline{\includegraphics[width=9cm]{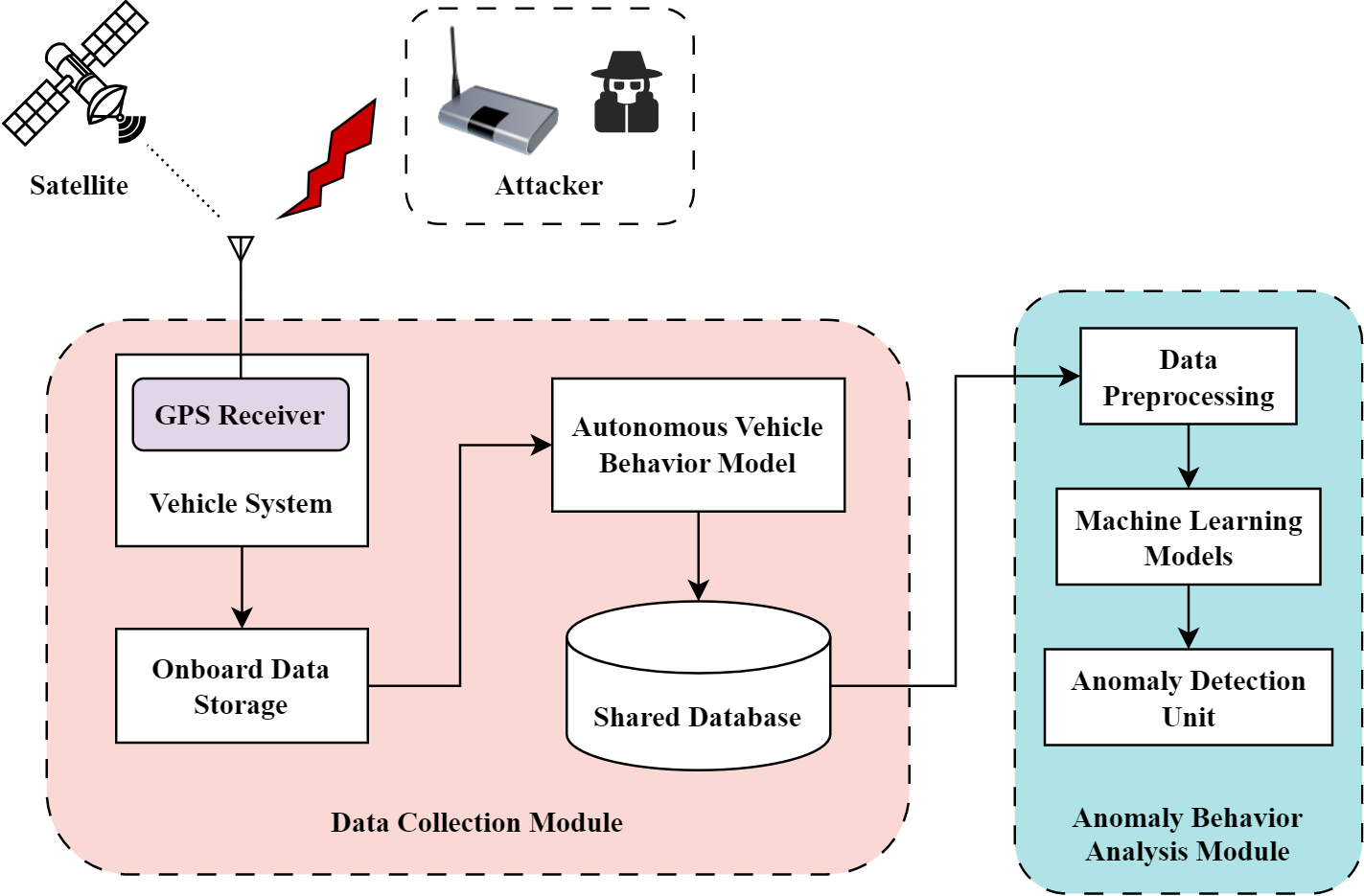}}  %[width=13cm]
\caption{GPS-IDS Architecture}
\label{gps ids architecture}
\end{figure}

\section{GPS Intrusion Detection System (GPS-IDS)}\label{gps ids} 

GPS-IDS is an Anomaly-based Intrusion Detection System designed to identify GPS spoofing attacks on AVs by continuously monitoring the behavior of the vehicle and considering any deviation from the expected behavior as anomalous. The framework relies on a hybrid vehicle model called the \textit{Autonomous Vehicle Behavior Model} to represent the vehicle's behavior using both physics-based and data-centric components at any instantaneous time $t$. Features highlighted in the physics-based component are used in machine learning analysis to detect anomalous behavior of the vehicle. Fig. \ref{gps ids architecture} shows the architecture of the GPS-IDS framework. The framework has two modules: the \textit{Data Collection Module}, and the \textit{Anomaly Behavior Analysis Module}.

\subsection{Data Collection Module} 

This module collects the raw data from the AV system and stores it in a shared database. It ensures that the necessary data is collected in real-time and is available for subsequent analysis. It has 3 main components: 1) Vehicle System, 2) Onboard Data Storage, and 3) Autonomous Vehicle Behavior Model.

% Raw data includes GPS signals and an ongoing stream of vehicle dynamics data, which are obtained from the onboard state-measuring sensors.
%authentic navigation data from satellites as well as malicious data from unauthorized attackers. The Autonomous Vehicle Behavior Model highlights the essential parameters necessary for estimating the current and future states of the vehicle, thereby offering guidance on the specific data parameters to be collected for the experimental analysis. Moreover, it serves the additional purpose of assessing the impact of GPS spoofing attacks on the vehicle. The shared database enables the sharing and transferring of information between the Data Collection Module and the Anomaly Behavior Analysis Module.

\subsubsection{Vehicle System and GPS Receiver}

The vehicle system includes a GPS receiver that is susceptible to spoofing attacks, where an attacker can manipulate the GPS receiver to provide false location data to the vehicle.

\subsubsection{Onboard Data Storage}

The onboard data storage system of the vehicle temporarily stores the raw data collected by the vehicle system. Raw data includes GPS signals and an ongoing stream of vehicle dynamics data, which are obtained from the onboard state-measuring sensors. This storage acts as a buffer that holds data before being analyzed by the subsequent modules.

\subsubsection{Autonomous Vehicle Behavior Model}\label{autonomus vehicle behavior model}

\begin{figure}
\centerline{\includegraphics[width=7.5cm]{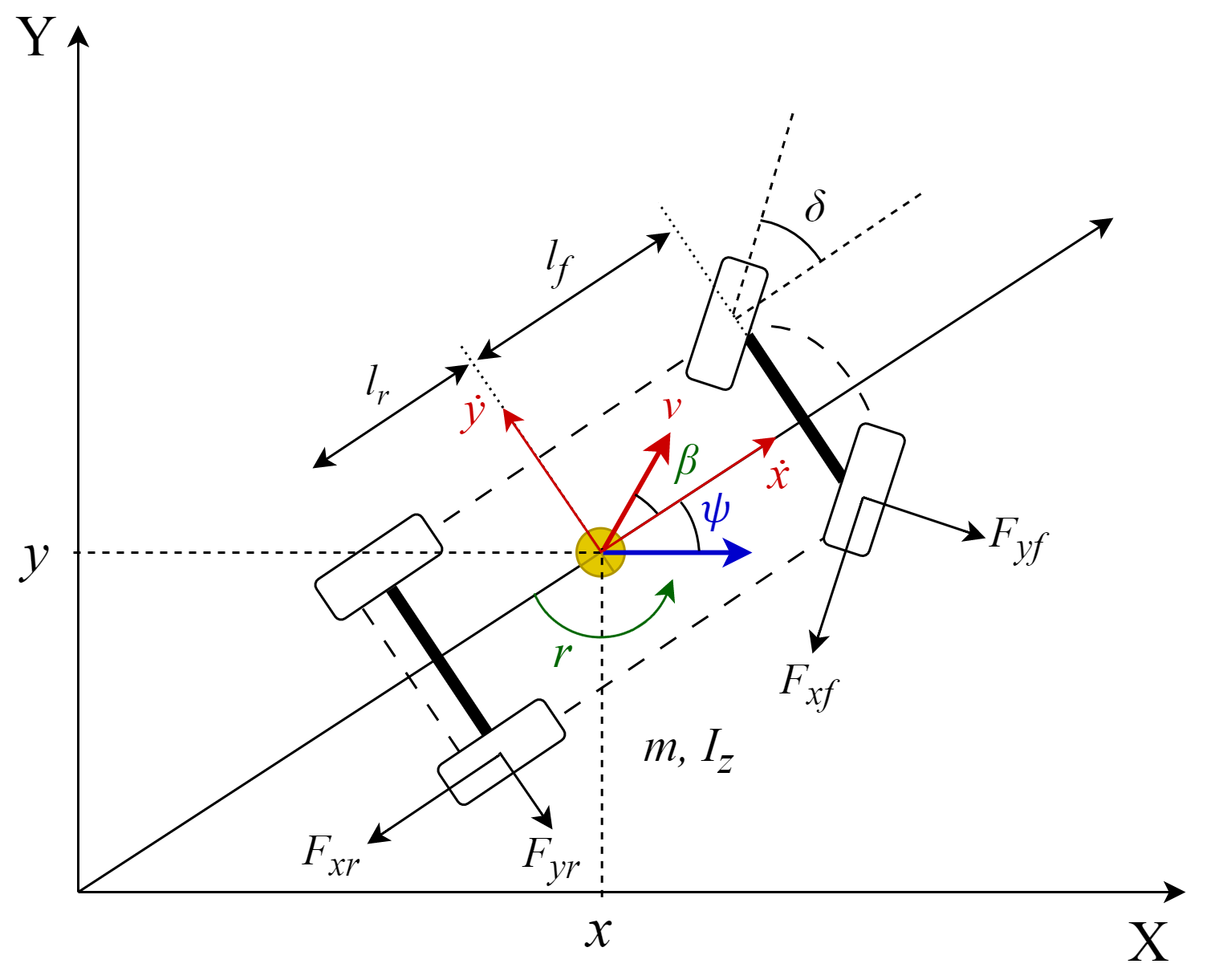}}  %[width=13cm]
\caption{Dynamic bicycle model of an autonomous vehicle in a 2-dimensional inertial frame}
\label{bicycle model}
\end{figure}

In order to establish a comprehensive behavior model of AVs, it is essential to consider four key components: (a) Localization or Navigation, (b) State Estimation, (c) Motion Planning, and (d) Control \cite{Perception planning control and coordination}--- and the Autonomous Vehicle Behavior Model is comprised of these four components. Localization relies on a combination of internal and external sensors (like GPS, IMUs, Cameras, LiDARs, etc.) to collect information about the surrounding environment. The remaining three components require a mathematical representation of the vehicle that encompasses its dynamics. This study focuses specifically on GPS-guided localization in AVs, aiming to investigate the impact of a spoofing attack on the vehicle's dynamics and motion. Therefore, predicting the dynamic state of the vehicle becomes crucial. To achieve this, a simple 2 Degrees of Freedom (2 DOF) dynamic bicycle model proposed by Rajamani et al. \cite{Vehicle dynamics and control} is employed. Fig. \ref{bicycle model} depicts the dynamic bicycle model of a vehicle in a 2-dimensional inertial frame.

%\subsubsection{Kinematic Bicycle Model}

%A kinematic bicycle model is a simplified representation of a vehicle's kinematics. This model mathematically describes the motion of a vehicle ignoring the effects of forces and moments acting on it. The following are the nonlinear continuous-time equations that describe a kinematic bicycle model in a 2-dimensional inertial frame- 

%\begin{equation}\label{eq1}
%v_x  = \dot{x} = v \cos(\psi + \beta)
%\end{equation}
%\begin{equation}\label{eq2}
%v_y = \dot{y} = v \sin(\psi + \beta)
%\end{equation}
%\begin{equation}\label{eq3}
%r = \dot \psi = \frac{v} {l_r} \sin{\beta}
%\end{equation}
%\begin{equation}\label{eq4}
%\dot v = a
%\end{equation}
%\begin{equation}\label{eq5}
%\beta = \tan^{-1} (\frac{l_r} {l_f + l_r} %\tan{\delta})
%\end{equation}

%where \(x\) and \(y\) are the coordinates of the center of mass of the vehicle in inertial frame (X, Y); \(v_x\) and \(v_y\) represent the longitudinal and lateral velocities of the vehicle, respectively; \(\psi\) is the yaw angle or heading angle, which represents the orientation of the vehicle with respect to the x-axis; \(r\) is the yaw rate, which is the rate of change of the yaw angle; \(a\) is the acceleration of the center of mass of the vehicle; \(\beta\) is the angle of the current velocity of the center of mass with respect to the longitudinal axis of the vehicle; \(\delta\) is the controlled input of the steering angle of the front wheels; and \(l_f\) and \(l_r\) are the distances from the center of mass to the front wheel axle and the rear wheel axle, respectively.

\paragraph{Dynamic Bicycle Model}\label{section dynamic bicycle model}

The dynamic bicycle model is a simplified representation of a vehicle's dynamics and considers the effects of external forces and yaw moments acting on the vehicle, which results in an accurate calculation of dynamic parameters \cite{Vehicle dynamics and control}. In a dynamic bicycle model, the inertial position coordinates and the orientation of the vehicle are defined as follows \cite{Kinematic and dynamic vehicle models for autonomous driving control design}:
\begin{equation}\label{eq1}
v_x  = \dot{x} = v \cos(\psi + \beta)
\end{equation}
\begin{equation}\label{eq2}
v_y = \dot{y} = v \sin(\psi + \beta)
\end{equation}
\begin{equation}\label{eq3}
r = \dot \psi = \frac{v} {l_r} \sin{\beta}
\end{equation}

where \(x\) and \(y\) are the coordinates of the center of mass of the vehicle in frame (X, Y); \(v_x\) and \(v_y\) represent the longitudinal and lateral velocities of the vehicle, respectively; \(\psi\) is the yaw angle, which is the orientation of the vehicle with respect to the x-axis; \(r\) is the yaw rate or the rate of change of the yaw angle; \(\beta\) is the angle of the current velocity of the center of mass with respect to the longitudinal axis of the vehicle; \(\delta\) is the controlled steering angle of the front wheels; and \(l_f\) and \(l_r\) are the distances of the front and the rear wheel axles from the center of mass, respectively. The differential equations associated with the dynamic bicycle model are:
\noindent
\begin{equation}\label{eq6}
\ddot{x} = \dot{\psi} \dot{y} + a_x
\end{equation}
\noindent
\begin{equation}\label{eq7}
\ddot{y} = - \dot{\psi} \dot{x} + \frac{2}{m} (F_{yf} \cos{\delta} + F_{yr})
\end{equation}
\noindent
\begin{equation}\label{eq8}
\dot{r} = {\ddot{\psi}} = \frac{2}{I_z} (l_f F_{yf} - l_r F_{yr})
\end{equation}
\noindent
%\begin{multicols}{2}
%\noindent
%\begin{equation}\label{eq9}
%\dot{X} = \dot{x} \cos{\psi} - \dot{y} \sin{\psi}
%\end{equation}
%\noindent
%\begin{equation}\label{eq10}
%\dot{Y} = \dot{x} \sin{\psi} - \dot{y} \cos{\psi}
%\end{equation}
%\end{multicols}

where $\dot{x}$ and $\dot{y}$ denote the longitudinal and lateral velocities of the vehicle, respectively; $a_x$ is the acceleration of the center of the mass; $\dot{\psi}$ or $r$ is the yaw rate; $m$ and $I_z$ denote the vehicle's mass and yaw inertia, respectively; and $F_{yf}$ and $F_{yr}$ denote the lateral tire forces at the front and rear wheels of the vehicle, respectively. From the dynamic bicycle model, Newton-Euler's equations of motion are defined as follows: 
\begin{equation}
\begin{bmatrix}
\boldsymbol{F}_x \\ 
\boldsymbol{F}_y
\end{bmatrix} 
= m
\begin{bmatrix}
\dot{v_x} - \dot{\psi} v_y \\
\dot{v_y} - \dot{\psi} v_x
\end{bmatrix}
= 
\begin{bmatrix}
-F_{xf} \cos{\delta} - F_{yf} \sin{\delta} - F_{xr} \\
F_{yf} \cos{\delta} - F_{xf} \sin{\delta} + F_{yr}
\end{bmatrix}
\end{equation}
\begin{equation} 
\boldsymbol{\tau} = I_z \ddot{\psi} = I_z \dot{r} = l_f (F_{yf} \cos{\delta} - F_{xf} \sin{\delta}) - l_r F_{yr}
\end{equation}

These dynamics can be simplified by disregarding the aerodynamic resistance and setting the longitudinal tire forces, \(F_{xf}\) and \(F_{xr}\), to zero. The lateral tire forces, \(F_{yf}\) and \(F_{yr}\) are calculated by using a linear tire model, which simplifies the nonlinear characteristics of tire dynamics by establishing a linear relationship between the tire slip angles and the resulting tire forces. Considering that, \(F_{yf}\) and \(F_{yr}\) are defined as \cite{Path planning using a dynamic vehicle model}:

\begin{multicols}{2}
\noindent
\begin{equation} 
F_{yf} = - C_{yf} \alpha_f 
\end{equation}%\break
\noindent
\begin{equation}
F_{yr} = - C_{yr} \alpha_r
\end{equation}
\end{multicols}

where \(C_{yf}\) and \(C_{yr}\) are the cornering stiffness coefficients, and \(\alpha_f\) and \(\alpha_r\) are the slip angles of the front and rear wheels, respectively. Assuming small slip angles, we obtain \cite{Path planning using a dynamic vehicle model}:

\begin{multicols}{2}
\noindent
\begin{equation}
\alpha_f = \frac{v_y + l_f r}{v_x} - \delta
\end{equation}%\break
\noindent
\begin{equation}
\alpha_r = \frac{v_y - l_r r}{v_x}
\end{equation}
\end{multicols}

Finally, a simplified 2 DOF non-linear state space representation of the lateral dynamics of the vehicle can be expressed as follows \cite{road vehicle state estimation, control of car-like robot}: 
\begin{equation}\label{eq17}
\begin{bmatrix}
\dot{v_y} \\
\dot{r}
\end{bmatrix} 
= 
\begin{bmatrix}
\textbf{a}_{11} & \textbf{a}_{12} \\
\textbf{a}_{21} & \textbf{a}_{22}
\end{bmatrix}
\begin{bmatrix}
v_y \\
r
\end{bmatrix}
+
\begin{bmatrix}
\textbf{b}_{11} \\
\textbf{b}_{21}
\end{bmatrix}
\delta
\end{equation}
\noindent
where,
\begin{multicols}{2}
\noindent
\begin{equation*}
\textbf{a}_{11} = \frac{C_{yf} + C_{yr}}{m v_x}
\end{equation*}%\break
\noindent
\begin{equation*}
\textbf{a}_{12} = \frac{l_f C_{yf} - l_r C_{yr}}{m v^2_x}
\end{equation*}
\end{multicols}
\begin{multicols}{2}
\noindent
\begin{equation*}
\textbf{a}_{21} = \frac{l_f C_{yf} - l_r C_{yr}}{I_z}
\end{equation*}%\break
\noindent
\begin{equation*}
\textbf{a}_{22} = \frac{l^2_f C_{yf} - l^2_r C_{yr}}{I_z v_x} 
\end{equation*}
\end{multicols}
\begin{multicols}{2}
\noindent
\begin{equation*}
\textbf{b}_{11} = - \frac{C_{yf}}{m v_x}
\end{equation*}%\break
\noindent
\begin{equation*}
\textbf{b}_{21} = - \frac{l_f C_{yf}}{I_z} 
\end{equation*}
\end{multicols}

In the state space model of equation \ref{eq17}, the input is the steering angle $\delta$ and the states are the lateral velocity $v_y$ and yaw rate $r$. 

\paragraph{Localization/ Navigation}

The GPS enables the vehicle to determine its position and localize to the destination. Since this paper focuses on presenting a GPS intrusion detection system, only GPS-based navigation is considered, and camera or vision sensor-based perception is not taken into account. To ensure safe GPS-guided localization for autonomous vehicles, it is necessary to continuously monitor various parameters such as GPS latitude and longitude, signal quality, GPS Dilution of Precision (DOP), and the number of satellites the vehicle is locked with. The normal behavior of the GPS signal for vehicular localization can be modeled by constantly monitoring the following parameters:

\begin{equation}\label{eq18}
s_{gps}(t) = (lat, lon, dop, sat_{lock}, sat_{count})^t
\end{equation}

where $s_{gps}(t)$ is the incoming GPS signal from satellites at time $t$; $lat$ and $lon$ are the GPS latitude and longitude at $t$, respectively; $dop$ denotes the Dilution of Precision (DOP) or the quality of the incoming GPS signal at $t$; and $sat_{lock}$ and $sat_{count}$ respectively denote the number of satellites the vehicle is locked with and the number of available satellites at $t$. It is noteworthy that the current GPS-guided localization behavior model only considers positional parameters and signal strength, excluding the analysis of physical layer parameters of GPS signal due to existing research on such GPS attack detectors, as outlined in section \ref{gnss security}.

\paragraph{State Estimation}

One of the well-established solutions for estimating the state of nonlinear systems is the Extended Kalman Filter or EKF. 
It integrates measurements from multiple sensors with a system model to estimate the state of the system with improved accuracy \cite{dynamics estimation ekf}. In the context of a GPS-guided AV, the EKF utilizes multiple sensor measurements in conjunction with the dynamic vehicle model to make accurate estimation about the vehicle's state, including its position, orientation, velocity, etc. This estimation procedure takes into account the nonlinear relationship between the measurements and the position, while also considering measurement noise. %As highlighted in the Related Work section, 
EKF state estimator with GPS/INS fusion is capable of detecting sensor anomalies and attacks against autonomous vehicles by monitoring the measurement deviations of multi-sensor readings \cite{cyberattack dynamic state estimation, secure pose estimation}. % We have employed a comparable EKF dynamic state estimator in our vehicle model. In later sections, it will be shown that our approach is able to detect a GPS spoofing attack faster than the employed EKF state estimation-based detection approach, which is supported by experimental validation.
A similar GPS/INS fusion using EKF dynamic state estimator has been employed in our vehicle model and experimental testbed. In Section \ref{experiments}, it is shown that the GPS-IDS approach is able to detect a GPS spoofing attack faster than the employed EKF-based GPS/INS detection approach, which is supported by experimental validation.

For ease of description, we can rewrite equation \ref{eq17} in the following forms:

\noindent
\begin{equation}\label{eq19}
\left.
\begin{aligned}
&\dot{\textbf{x}}(t) = f_{state}(\textbf{x}(t), u(t))\\
&\dot{\textbf{x}} = \textbf{ax} + \textbf{b} u\\
\end{aligned}
\right\}
\end{equation}

where vehicle state $\textbf{x} = [v_y \hspace{0.2cm} r]^t$, control input $u = [\delta]^t$, and $f_{state}$ is the nonlinear function reproducing equation \ref{eq17}. According to the dynamic bicycle model, the normal behavior of the vehicle's lateral dynamics is characterized by the state parameters, $v_y$ and $r$, and the steering angle $\delta$. In conjunction with the lateral dynamics model, the EKF allows us to fuse and compare multiple state-measuring sensor readings to estimate the vehicle's state with improved accuracy. To incorporate an EKF-based estimation, the vehicle can be described as a discrete time-varying system in the following forms:
\noindent
%\begin{multicols}{2}
\noindent
\begin{equation}\label{se1}
    \textbf{x}_{k+1} = f_{cd}(\textbf{x}_k, u_k, W_k) 
\end{equation}
\noindent
\begin{equation}\label{se2}
    \textbf{y}_{k+1} = g_{cd}(\textbf{x}_k, E_k) 
\end{equation}
%\end{multicols}

where $f_{cd}$ is the prediction equation; $\textbf{x}_k$ and $u_k$ are the state and the input at the $k^{th}$ time, respectively; $W_k$ is the prediction noise; $g_{cd}$ is the observation equation; and $E_k$ denotes the observation noise. The Jacobian matrix of the nonlinear prediction and observation equation are defined as: 

\begin{multicols}{2}
\noindent
\begin{equation}\label{f}
    F = \left.\frac{\partial f_{cd}}{\partial \textbf{x}}\right|_{\hat x_{k-1}, u_k}
\end{equation}%\break
\noindent
\begin{equation}\label{g}
    G = \left.\frac{\partial g_{cd}}{\partial \textbf{x}}\right|_{\hat x_{k}}
\end{equation}
\end{multicols}

Thus, the nonlinear system can be transformed into a linear system with an update process by the following equations:

\begin{multicols}{2}
\noindent
\begin{equation}\label{senonlinear1}
    \textbf{x}_{k+1} = F \textbf{x}_k + W_k 
\end{equation}
\noindent
\begin{equation}\label{senonlinear2}
    \textbf{y}_{k+1} = G \textbf{x}_k + E_k
\end{equation}
\end{multicols}
%The update process is as follows:
%\begin{equation*}
%    \textbf{P}_{\left.{k+1} \right|{k}} = F_k \textbf{P}_{\left.k \right| k} F^T_k + \textbf{Q}
%\end{equation*}
%\begin{equation*}
%    \textbf{S}_{\left.{k+1} \right|{k}} = G_k \textbf{P}_{\left.{k+1} \right| k} G^T_k + \textbf{R}
%\end{equation*}
%\begin{equation*}
%    \textbf{K}_{\left.{k+1} \right|{k}} = \frac{\textbf{P}_{\left.{k+1} \right|k} G^T_k}{\textbf{S}_{\left.{k+1} \right|k}}
%\end{equation*}
%\begin{equation*}
%    \textbf{P}_{\left.{k+1} \right|{k+1}} = (\textbf{I} - \textbf{K}_{\left.{k+1} \right| k} G_k) \textbf{P}_{\left. {k+1} \right| k}
%\end{equation*}

%where $\textbf{Q}$ is the covariance matrix of the prediction noise, $\textbf{R}$ is the covariance matrix of the observation noise, and \textbf{I} is an identity matrix. Thus: 
\begin{multicols}{2}
\noindent
\begin{equation}\label{seupdate1}
    \textbf{x}_{\left.{k+1} \right|k} = F \hat{\textbf{x}}_{\left.k \right|k} + W_k 
\end{equation}
\noindent
\begin{equation}\label{seupdate2}
    \textbf{y}_{\left.{k+1} \right|k} = G \hat{\textbf{x}}_{\left.k \right|k} + E_k
\end{equation}
\end{multicols}

The final EKF estimated value can be obtained by: 
\begin{equation}\label{seekf}
    \hat{\textbf{x}}_{\left.{k+1} \right|{k+1}} = x_{\left.{k+1} \right|k} + \textbf{K}_{\left.{k+1} \right|k} (\textbf{y}_{k+1} - \textbf{y}_{\left.{k+1} \right|k})
\end{equation}

\paragraph{Motion Planning}
The vehicle plans a safe and efficient path to follow from its current position $(x, y)$ to the target destination $(x_t, y_t)$. It uses GPS to determine its current position and generates a continuous sequence of target yaw angles, $\psi_t(t)$ and cross-track errors, $e(t)$ to guide itself to the desired path. The cross-track error refers to the perpendicular distance of the vehicle from the current position to the desired path. Algorithm \ref{target angle algorithm} outlines the target yaw angle calculation process from GPS coordinates \cite{An autonomous delivery robot to prevent the spread of coronavirus}.

\begin{algorithm}[H]
\caption{Target Yaw Calculation from GPS Coordinates}\label{target angle algorithm}
%\hspace*\algorithmicindent{\textbf{Input:}} {Current Longitude = $lon_c$}\\
%\hspace*\algorithmicindent{Current Latitude = $lat_c$}\\
%\hspace*\algorithmicindent{Current Longitude = $lon_c$}\\
%\hspace*\algorithmicindent{Target Latitude = $lat_t$}\\
%\hspace*\algorithmicindent{Target Longitude = $lon_t$}\\
%\hspace*\algorithmicindent{\textbf{Ouput:}} {Target Yaw Angle $\psi_t$}\\
\begin{algorithmic}[1]
\STATE \textbf{Function} $TargetYawFromGPS$ (current\_latitude, \hspace{1cm} current\_longitude, target\_latitude, target\_longitude) 
\WHILE{($TargetYawFromGPS$ = True)}
\STATE $\Delta$ longitude $\gets$ %radians(
target\_longitude -- current\_longitude\\
%\STATE current\_latitude $\gets$ radians(current\_latitude)\\
%\STATE target\_latitude $\gets$ radians(target\_latitude)\\
\STATE $p = \sin$ ($\Delta$ longitude) $\times \cos$ (target\_latitude)\\
\STATE $q = \cos$ (current\_latitude) $\times$ $\sin$ (target\_latitude) \\ \hspace{0.19cm} -- $\sin$ (current\_latitude) $\times \cos$ (target\_latitude) \\ \hspace{0.19cm} $\times \cos$ ($\Delta$ longitude)\\
\STATE $\psi_t = arctan(p, q)$ \hspace{1.6cm} [$\psi_t$ = target yaw angle]\\
\STATE \textbf{return} \hspace{0.1cm} $\psi_t$ 
\ENDWHILE
\end{algorithmic}
\end{algorithm}

From the calculated target yaw, $\psi_t(t)$ and current yaw, $\psi(t)$, the motion planning algorithm can calculate the instantaneous cross-track error, $e(t)$ using the following relationship: %\(e(t) \propto \sin{(\theta_T(t) - \theta(t))}\)

\begin{equation}\label{eq20}
e(t) \propto \sin{(\psi_t(t) - \psi(t))}
\end{equation}

From \ref{eq20}, we can say that when the vehicle is traveling along the desired path, $\psi(t) = \psi_t(t)$, and $e(t) = 0$. When the vehicle deviates from the desired path, $\psi(t)$ differs from $\psi_t(t)$ and the $e(t)$ increases accordingly. In general, an increase in the difference between the target yaw angle and the current yaw angle leads to a corresponding amplification in the cross-track error.

\paragraph{Controller}

The vehicle requires a control algorithm to mitigate the cross-track error and navigate safely. There are different types of controllers depending on vehicle-specific models and computational resources, such as Model Predictive Controller (MPC), Fuzzy Logic Controller, Proportional-Integral-Differential (PID) Controller, etc. In this study, we focus on the PID controller as it is one of the most widely used control algorithms and easy to implement on testbeds.

\begin{figure}[t]
\centerline{\includegraphics[width=9cm]{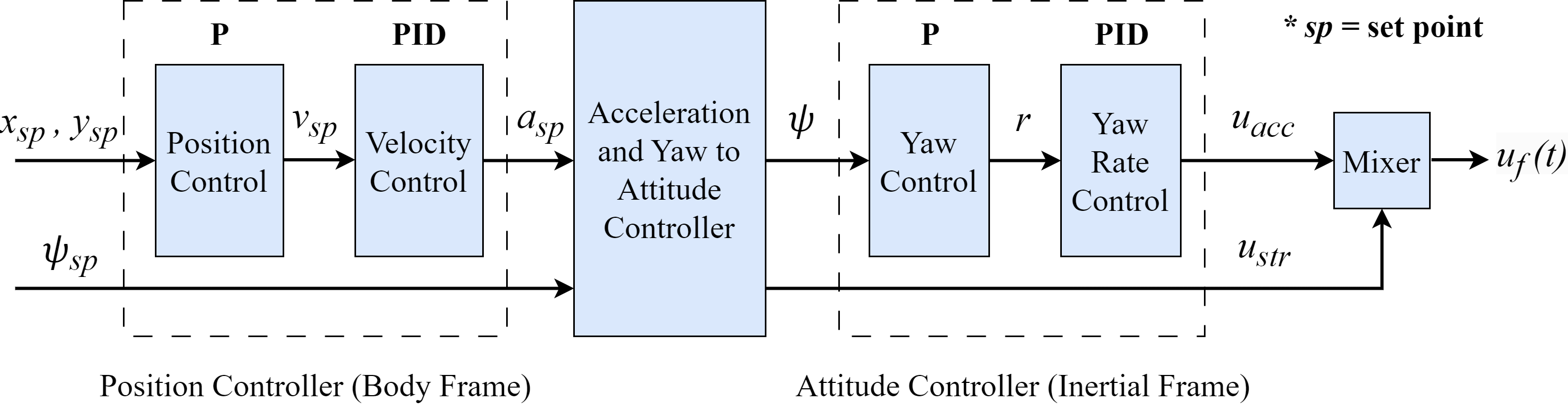}}  %[width=13cm]
\caption{PID Cascade Control Architecture}
\label{PID}
\end{figure}

The PID controller is typically used for controlling the \textit{Position} (e.g. latitude and longitude) and \textit{Attitude} (e.g. yaw/ heading angle) of an AV. The controller continuously adjusts the steering and acceleration control signals to minimize the cross-track error. According to the proportional (P), integral (I), and derivative (D) terms of $e(t)$, the control signal $u_c(t)$ is obtained, which is formulated as follows:

\noindent
\begin{equation}\label{eq21}
u_c(t) = K_p e(t) + K_i \int\limits_0^t e(t)dt + K_d \frac{d e(t)}{dt}
\end{equation}

where $K_p$, $K_i$, and $K_d$ are the Proportional, Integral, and Differential gain coefficients, respectively. The term $e(t)$ denotes the present cross-track error, $\int\limits_0^t e(t)dt$ denotes the accumulated cross-track error over time, represented as the definite integral of $e(t)$ with respect to time, and $\frac{d e(t)}{dt}$ is the change in cross-track error with respect to time, represented as the derivative of $e(t)$ with respect to time. The final control signal $u_f(t)$ will be comprised of an acceleration component and a steering angle component, which can be expressed as: 
%\noindent
\begin{equation}\label{eq22}
u_f(t) = u_{acc}(t) + u_{str}(t) = K_{acc} a(t) + K_{str} \delta(t)
\end{equation}

where $u_{acc}(t)$ and $u_{str}(t)$ are the acceleration and steering control signal components, respectively. $K_{acc}$ is the proportional gain for acceleration, $K_{str}$ is the proportional gain for steering, and $a(t)$ and $\delta(t)$ are instantaneous acceleration and steering angle, respectively. Fig. \ref{PID} illustrates the PID cascade control architecture used to control the position and attitude of the AV model.

\begin{figure*}[htbp]
\centerline{\includegraphics[width=16cm]{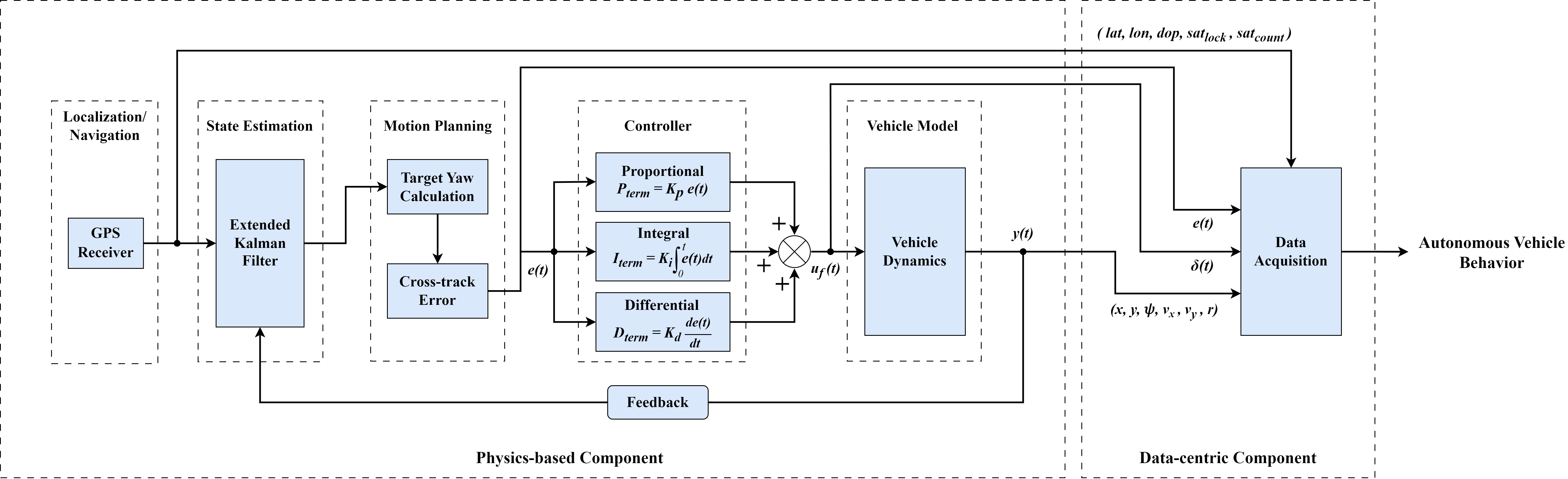}} %[width=13cm]
\caption{Autonomous Vehicle Behavior Model}
\label{Autonomous Vehicle Behavior Model}
\end{figure*}

Derived from the discussed software elements, the architecture of the Autonomous Vehicle Behavior Model is depicted in Fig. \ref{Autonomous Vehicle Behavior Model}. From the figure, we can observe that the model is established on 14 features--- Localization/ Navigation block: Latitude ($lat$), Longitude ($lon$), Horizontal DOP ($hdop$), Vertical DOP ($vdop$), Satellite Lock ($Sat_{lock}$), Satellite Count ($Sat_{count}$); Motion Planning block: Cross-Track Error ($e$); Controller block: Steering Angle ($\delta$); and Vehicle Dynamics block: Longitudinal Position ($x$), Lateral Position ($y$), Yaw Angle ($\psi$), Longitudinal Velocity ($v_x$), Lateral Velocity ($v_y$), and Yaw Rate ($r$). Data from these features are acquired by the Data-centric Component, and machine learning techniques are employed to model AV behavior and detect anomalies. 

\subsection{Anomaly Behavior Analysis (ABA) Module}

%This module reads the raw data from the shared database and performs Anomaly Behavior Analysis to classify the data as normal or abnormal. 

%\subsection{Anomaly Behavior Analysis (ABA) for GPS-IDS}

Satam et al. have presented an IDS for wireless networks based on Anomaly Behavior Analysis (ABA) \cite{WIDS, Hamid Alipour}, which forms the basis of this module. However, we improved it by integrating both physics and data, rather than relying solely on data to detect GPS spoofing on AVs. 

The ABA module for AVs is defined over a finite set of driving events $U$. Set $U$ is partitioned into two subsets: $Normal$ events $N$ and $Abnormal$ or $Attack$ events $A$, such that $N \cup A = U$ and $N \cap A = \emptyset$. %To characterize $U$, a representation map $R$ is used,
A representation map $R$ maps events in $U$ to patterns in $U^R$ such that $U \xrightarrow{R} U^R$. Likewise, $N^R$ and $A^R$ respectively represent the events in $N$ and $A$, such that $N \xrightarrow{R} N^R$, $A \xrightarrow{R} A^R$, and $N^R \cup A^R = U^R$. A detector $D$ is defined as $D = (f_{norm}, M)$; where $f_{norm}$ is the normal behavior characterization function expressed as $f_{norm}: U^R \times M \Rightarrow [0, 1]$ and $M$ is the system memory that stores the normal behavior model extracted from the set of normal events $N^R$. Function $f_{norm}$ specifies the degree of abnormality of a sample $s \in U^R$ by comparing it with $M$. The higher the value of $f_{norm}(s, M)$, the more abnormal the sample is. If the value of $f_{norm}(s, M)$ exceeds a predefined threshold $\mathbb{T}$, detector $D$ raises an alarm indicating the occurrence of an abnormal or attack event. We can consider $D$ for any sample $s \in U^R$ as:
\noindent
\begin{equation}
D(s) = 
\begin{cases}
    Abnormal \hspace{1cm} if \quad f_{norm}(s, M) > \mathbb{T}\\
    Normal \hspace{1.3cm} otherwise
\end{cases}
\end{equation}

%\begin{figure}[htbp]
%\centerline{\includegraphics[width=8cm]%{AnomalyDiagram.png}}  %[width=13cm]
%\caption{Anomaly Behavior Analysis %Methodology \cite{bib9}}
%\label{fig0}
%\end{figure}

%Detection takes place when the detector $D$ classifies a sample as abnormal, regardless of whether it is genuinely an anomaly or a regular sample that has been wrongly classified as one. 
Detection occurs when detector $D$ identifies a sample as abnormal, whether it is truly an anomaly or a normal sample misclassified as one. The detection errors are defined over a test set $U_t^R$ which is a subset of $U^R$, $U_t^R \subseteq U^R$. The detector considers two kinds of errors: \textit{False Positives} and \textit{False Negatives}. A \textit{False Positive} detection occurs when a normal sample $s \in N^R$ is detected as an abnormal event and is defined as $\varepsilon^+ = \{s \in N^R | D(s) = abnormal\}$, while a \textit{False Negative} detection occurs when the detector classifies an abnormal sample $s \in A^R$ as a normal event (undetected anomalies), that is $\varepsilon^- = \{s \in A^R | D(s) = normal\}$. The objective of GPS-IDS is to tune the predefined threshold $\mathbb{T}$, so that the overall error is minimized. Particularly, we prioritize the minimization of \textit{False Negative} errors, as these undetected attacks pose greater risks compared to false alarms in the context of AVs. In the GPS-IDS approach, ABA is applied using machine learning (Experiment 3: \ref{ABA_exp3}) and the threshold $\mathbb{T}$ is tuned from optimization (Experiment 4: \ref{threshold_tuning}).

%\subsection{AVT system model}

%The combined architecture of the AVT system is depicted in Figure \ref{fig4}. It comprises of a GPS receiver unit with an Extended Kalman Filter, a PID control unit, and a vehicle model unit that merges vehicle dynamics and kinematics. The AVT implements a feedback control mechanism that is used to calculate the cross track error \(e(t)\). The final output, \(y(t)\) is a controlled motion output which is a function of the current dynamic state, \([v_y \quad r]\).

\section{Attacker Model and Problem Statement}\label{attacker model}
%\textcolor{red}{ABRAR, MODIFY THIS SECTION TO INCLUDE THE STEALTHY PIECE OF ATTACK WE TALKED OF YESTERDAY. THE MODEL MUST READ IN SUCH A WAY THAT IT INCLUDES BOTH STEALTHY SPOOFING AND NON-STEALTHY SPOOFING}
We focus on the GPS spoofing attack that interferes with the target AV by injecting a malicious GPS signal into the vehicle's GPS receiver unit. The attacker is modeled as a malicious entity capable of generating, recording, and transmitting fake GPS signals with false coordinates corresponding to any location of his choice wirelessly using a spoofer device. The attacker can be an internal entity who is inside the victim vehicle and very close to the GPS receiver, or an external entity who is moving along with the victim vehicle with a directional antenna pointed toward it and staying within a specified range. It is assumed that the attacker only knows the current location of the victim vehicle and has no knowledge about (1) the low-level control algorithm settings; (2) control commands from the autonomous navigation system of the vehicle. In this study, it is shown that even though the attacker has no prior knowledge about the vehicle's control system, a GPS spoofing attack is capable of compromising the closed-loop control system and the measured states of the vehicle. 

\subsection{Mathematical model of the spoofed GPS signal} 

To model a spoofed GPS signal, the attacker must replicate the Radio Frequency carrier wave, the Pseudorandom noise code (PRN), and the data bits of each GPS signal that he or she intends to spoof. As stated by Psiaki and Humphreys, \cite{gnss spoofing and detection} a typical GPS signal $y(t)$ takes the following form:

\noindent
\begin{equation}%\label{eq23}
y(t) = \textbf{Re} \Bigl\{ \sum_{i=1}^{N} A_i D_i [t-\tau_i(t)] C_i[t-\tau_i(t)]e^{j \varphi} \Bigl\}
\end{equation}
\noindent
\begin{equation}
\varphi = \omega_c t - \phi_i(t)
\end{equation}

where $N$ is the number of individual signals transmitted by each GPS satellite. $A_i$, $D_i(t)$, and $C_i(t)$ correspond to the carrier amplitude, data bit stream, and spreading code (often a Binary Phase Shift Keying, BPSK-PRN code or a Binary Offset Carrier, BOC-PRN code) of the $i$th signal, respectively. $\tau_i(t)$ is the $i$th signal’s code phase, $\omega_c$ is the nominal carrier frequency, and $\phi_i(t)$ is the $i$th beat carrier phase. The attacker sends a set of spoofed signals $y_{spf}(t)$ that are similar to as follows:

\noindent
\begin{equation} %\label{eq24}
y_{spf}(t) = \textbf{Re} \Bigl\{ \sum_{i=1}^{N_{spf}} A_{i_{spf}} \hat{D_i} [t-\tau_{i_{spf}}(t)]C_i[t-\tau_{i_{spf}}(t)]e^{j \hat{\varphi}} \Bigl\}
\end{equation}
\noindent
\begin{equation}
\hat{\varphi} = \omega_c t - \phi_{i_{spf}}(t)
\end{equation}

where $N_{spf}$ is the number of spoofed signals (typically $N_{spf} = N$). Each spoofed signal must have the same spreading code $C_i(t)$ as the corresponding true signal in order to deceive the receiver, and usually, it broadcasts its best estimate of the same data bit stream $\hat{D}_i(t)$. The spoofed amplitudes, code phases, and carrier phases are, respectively, $A_{i_{spf}}$, $\tau_{i_{spf}}(t)$, and $\phi_{i_{spf}}(t)$ for $i = 1, ..., N_{spf}$. According to the mathematical model, when a spoofing attack is successful, the victim vehicle's GPS receiver antenna receives $y(t) + y_{spf}(t)$.

This attack is still detectable by several approaches discussed in \ref{related work} section. To enhance the stealthiness of our GPS spoofing attack, realistic noise was added to the generated spoofed signals by sampling a Gaussian distribution with a large variance, as detailed by Wang et al. \cite{wang2023infra}. The addition of Gaussian noise aims to mimic the natural variations and imperfections present in genuine GPS signals, making it more challenging for IDSs to identify the spoofing attack.

Finally, the total signal received, $y_{total}(t)$ at the victim vehicle's GPS receiver antenna is:

\noindent
\begin{equation}\label{eq25}
y_{total}(t) = y(t) + y_{spf}(t) + \nu(t)
\end{equation}

where $\nu(t)$ is the Gaussian noise component added to the spoofing attack to make it stealthy.

%The attacker initiates the attack by transmitting a spoofed GPS signal with false GPS coordinates. The conditions for the attack to be successful can be expressed as: \(\langle f_{spf} = f_{gps}, P_{spf} > P_{gps} \rangle \); where \(f_{spf}\) and \(f_{gps}\) denote the frequency of the spoofed signal and the authentic GPS signal, respectively; and \(P_{spf}\) and \(P_{gps}\) denote the power of the spoofed signal and the authentic GPS signal, respectively. 

\subsection{Mathematical model of the attack impacts on the Vehicle}

If the attack is successful, the GPS receiver of the vehicle will receive incorrect location information, and the current latitude and longitude will be replaced by spoofed latitude and longitude. The function $TargetYawFromGPS()$ in Algorithm \ref{target angle algorithm} takes the spoofed latitude and longitude as inputs and calculates the spoofed yaw angle, \(\psi_{t_{spf}}\). This means that the proportionality of equation \ref{eq20} calculates a spoofed cross-track error, \(e_{spf}(t)\). The PID controller takes \(e_{spf}(t)\) as input and equation \ref{eq21} and \ref{eq22} become:

\noindent
\begin{equation}\label{eq26}
u_{spf}(t) = 
\begin{cases}
    K_p e_{spf}(t) + K_i \int\limits_0^t e_{spf}(t)dt + K_d \frac{d e_{spf}(t)}{dt} \\
    K_{acc} a_{spf}(t) + K_{str} \delta_{spf}(t) 
\end{cases}
\end{equation}

where $u_{spf}(t)$, $a_{spf}(t)$, and $\delta_{spf}(t)$ represent the spoofed control signal, spoofed acceleration, and spoofed steering angle, respectively. $u_{spf}(t)$ goes to the vehicle system as actuator command, which results in $\delta_{spf}(t)$ and the vehicle loses its control at a time instance $t$. The state space representation of equation \ref{eq17} thus becomes:
\begin{equation}\label{eq27}
\begin{bmatrix}
\dot{v_y}_{spf}\\
\dot{r}_{spf}
\end{bmatrix} 
= 
\begin{bmatrix}
\textbf{a}_{11} & \textbf{a}_{12} \\
\textbf{a}_{21} & \textbf{a}_{22}
\end{bmatrix}
\begin{bmatrix}
v_y \\
r
\end{bmatrix}
+
\begin{bmatrix}
\textbf{b}_{11} \\
\textbf{b}_{21}
\end{bmatrix}
\delta_{spf}
\end{equation}

%Where, 
%\begin{equation*}
%\bar{\textbf{a}}_{11} = \frac{C_f \cos{\delta_{spf}} + C_2}{m v_x}
%\end{equation*}
%\begin{equation*}
%\bar{\textbf{a}}_{12} = \frac{-l_f C_1 \cos{\delta_{spf}} + l_r C_2}{I_z v_x}
%\end{equation*}
%\begin{equation*}
%\bar{\textbf{a}}_{21} = \frac{-l_f C_1 \cos{\delta_{spf}} + l_r C_2}{m v_x} - v_x
%\end{equation*}
%\begin{equation*}
%\bar{\textbf{a}}_{22} = - \frac{l^2_f C_1 \cos{\delta_{spf}} + l^2_r C_2}{I_z v_x} 
%\end{equation*}
%\begin{equation*}
%\bar{\textbf{b}}_{11} = \frac{C_1 \cos{\delta_{spf}}}{m} 
%\end{equation*}
%\begin{equation*}
%\bar{\textbf{b}}_{21} = \frac{l_f C_1 \cos{\delta_{spf}}}{I_z} 
%\end{equation*}

where [$\dot{v_y}_{spf}$ $\dot{r}_{spf}$] is the spoofed lateral dynamic state of the vehicle. In this state, the vehicle will start to deviate from its normal lateral dynamic behavior and move toward the spoofed GPS location. Based on the discussion, the flow of the GPS spoofing attack is summarized in Algorithm \ref{spoofing algorithm}.
%\textcolor{red}{MAKE SURE THIS SPOOFING ATTACK ALOGRITHM (ALG2) CAPTURES THE RESTRICTIONS THAT MAKE IT STEALTHY}
\begin{algorithm}
\caption{GPS Spoofing Attack}\label{spoofing algorithm}
\hspace*\algorithmicindent{\textbf{Input:}} {Spoofed signal $y_{spf}$, Duration of attack $t_{attack}$,}\\
\hspace*\algorithmicindent{Bias angle caused by attack $\vartheta$, Current state $\textbf{x}$,}\\
\hspace*\algorithmicindent{Spoofed state $\textbf{x}_{spf}$, Gaussian noise component $\nu(t)$,}\\
\hspace*\algorithmicindent{current\_latitude, current\_longitude}\\
\hspace*\algorithmicindent{spoofed\_latitude, spoofed\_longitude}\\
\hspace*\algorithmicindent{\textbf{Output:} Attack flag \textbf{A}, State update $\textbf{x}_{new}$}

\begin{algorithmic}[1]
\STATE Attacker sends $y_{spf} + \nu \Rightarrow \text{stealthy attack}$

\IF{receiver antenna receives $y_{total} = y + y_{spf} + \nu$}
\STATE \textbf{A} = True \\
\ELSE \STATE \textbf{A} = False\\
\WHILE {($t_{attack}$ = True)}
\STATE Compute $\psi_{t_{spf}}$ using $TargetYawFromGPS$\\
\STATE (current\_longitude, current\_latitude, \\
\STATE spoofed\_longitude, spoofed\_latitude) \hspace{0.2cm} [algorithm \ref{target angle algorithm}]\\
\STATE $\psi_{t_{spf}} \gets \psi_t + \vartheta$ \hspace{1.8cm} [update $\psi_t$ with bias $\vartheta$]\\
\STATE $e_{spf} \Rightarrow \textbf{\textit{f}} (\psi_{t_{spf}} - \psi_t)$ \hspace{1.37cm} [using equation \ref{eq20}]
\STATE $u_{spf} \Rightarrow e_{spf}$ \hspace{2.625cm} [using equation \ref{eq26}]
\STATE $\delta_{spf} \Rightarrow u_{spf}$
\STATE Update \textbf{x} \hspace{3.2cm} [using equation \ref{eq19}]

\ENDWHILE
\ENDIF
\end{algorithmic}
\end{algorithm}

%\begin{algorithm}
%\caption{GPS Spoofing Attack}\label{spoofing algorithm}
%\hspace*\algorithmicindent{\textbf{Input:}} {Spoofed signal $y_{spf}$}\\
%\hspace*\algorithmicindent{Duration of attack $t_{attack}$}\\
%\hspace*\algorithmicindent{Bias angle caused by attack $\vartheta$}\\ 
%\hspace*\algorithmicindent{EKF threshold $Th_{ekf}$}\\
%\hspace*\algorithmicindent{Current state $\textbf{x}$, Spoofed state $\textbf{x}_{spf}$}\\ 
%\hspace*\algorithmicindent{current\_latitude, current\_longitude}\\
%\hspace*\algorithmicindent{spoofed\_latitude, spoofed\_longitude}\\
%\hspace*\algorithmicindent{\textbf{Output:} Attack flag \textbf{A}, State update $\textbf{x}_{new}$}\\
%\begin{algorithmic}[1]
%\STATE Attacker sends $y_{spf}$
%\IF{receiver antenna receives $y_{total} = y + y_{spf} + \nu$}
%\STATE \textbf{A} = True\\
%\ELSE \STATE \textbf{A} = False\\
%\WHILE {($t_{attack}$ = True)}
%\STATE Compute $\psi_{t_{spf}}$ using $TargetAngleFromGPS$\\
%\STATE (current\_longitude, current\_latitude, \\
%\STATE spoofed\_longitude, spoofed\_latitude) [using algorithm \ref{target angle algorithm}]\\
%\STATE $\psi_{t_{spf}} \gets \psi_t + \vartheta$ \hspace{1.8cm} [update $\psi_t$ with bias $\vartheta$]\\
%\STATE $e_{spf} \Rightarrow \textbf{\textit{f}} (\psi_{t_{spf}} - \psi_t)$ \hspace{1.35cm} [using equation \ref{eq20}]
%\STATE $u_{spf} \Rightarrow e_{spf}$ \hspace{2.62cm} [using equation \ref{eq26}]
%\STATE $\delta_{spf} \Rightarrow u_{spf}$
%\STATE Update \textbf{x} \hspace{3.2cm} [using equation \ref{eq19}]

%\ENDWHILE
%\ENDIF
%\end{algorithmic}
%\end{algorithm}

\subsection{Problem Statement}

In autonomous driving applications, GPS spoofing attacks involve the deliberate manipulation of GPS signals to deceive the vehicle's onboard navigation system, which can cause the vehicle to deviate from its intended path, leading to accidents or even hijacking. To ensure the safety of passengers and pedestrians, attacks must be detected in a timely manner. Based on the above discussions, the GPS intrusion detection problem investigated in this paper can be stated as follows:

\textit{Let us consider an AV modeled by Fig. \ref{Autonomous Vehicle Behavior Model}, that has a state space representation described by equation \ref{eq17} under normal conditions and equation \ref{eq27} under the influence of a GPS spoofing attack. Assuming the vehicle is under normal operating conditions at the initial time, we have to design an anomaly-based intrusion detection strategy to detect GPS anomalies and, in turn, detect the GPS attacks.}

\section{Experimental Evaluation}\label{experiments}

To validate the proposed GPS-IDS framework and demonstrate its effectiveness, field experiments were conducted using an AV robotic testbed. The testbed was designed and developed by following the Autonomous Vehicle Behavior Model presented in Fig. \ref{Autonomous Vehicle Behavior Model}. This testbed was utilized to perform GPS spoofing attack experiments and collect relevant real-world data. In addition, simulated data with urban environment scenarios were generated using CARLA simulator. %\AY{We also used CARLA simulator \cite{CARLA} to run simulations in an urban environment and collected 3 datasets.} 
All the collected datasets have been used to conduct a series of experiments and validate the GPS-IDS framework.

\subsection{Field Experiments and Data Collection}
\subsubsection{Autonomous Vehicle Testbed (AVT)}

The Autonomous Vehicle Testbed or AVT is a custom-built autonomous rover that can navigate through a predefined path using GPS guidance. %It is designed to follow a predefined path and successfully navigate to the intended destination. 
It utilizes an array of state-measuring sensors, including an accelerometer, magnetometer, barometer, gyroscope, and digital compass for accurate state estimation. Following the Autonomous Vehicle Behavior Model, the AVT employs a PID controller and a GPS/INS fusion-based EKF failsafe. The EKF failsafe is triggered when the EKF variances associated with any two state-measuring sensor readings exceed a predefined EKF threshold value for 1 second. The AVT uses Ardupilot \cite{ardupilot}, an open-source software and hardware platform designed for building custom unmanned ground and aerial robotic vehicles. An Ardupilot-based autopilot generates the velocity and steering angle commands from the GPS and feedback from the state-measuring sensors of the AVT. These commands are then passed to the PID controller for execution. To program the AVT and define a path to follow using GPS, a Ground Control Station software supported by Ardupilot was utilized. The Ground Control Station computer receives all vehicular data via the telemetry modules in real time and stores them in the local memory of the computer. These vehicular data are referred to as ``Dataflash logs", and collected at Ardupilot's default data update rate of 1 Hz.

\begin{figure*}%[htbp]
\centerline{\includegraphics[width=12.5cm]{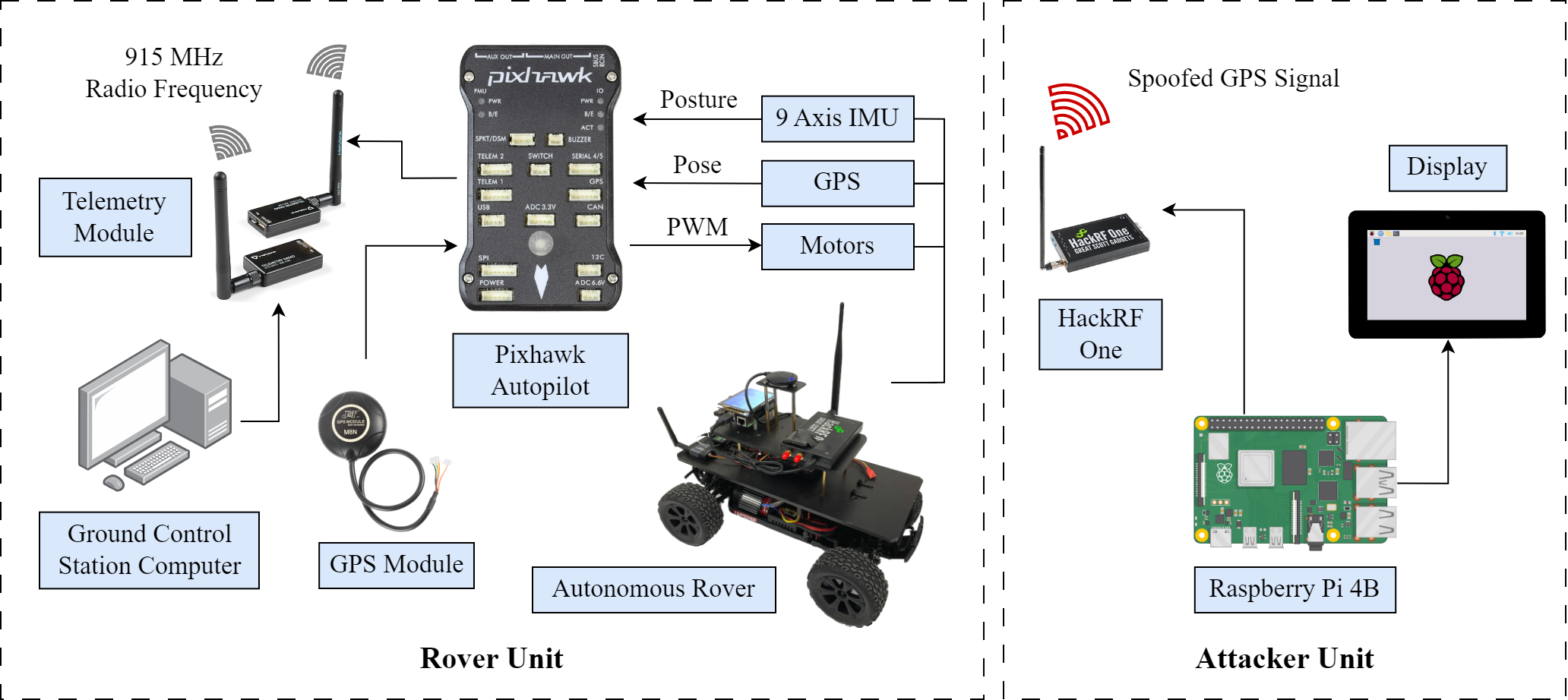}}
\caption{Hardware Architecture of the AVT}
\label{avt}
\end{figure*}

The hardware architecture of the AVT is divided into two parts: the Rover Unit and the Attacker Unit. The Rover Unit consists of a $1/10^{th}$ scale Radio Controlled truck chassis, a Pixhawk flight controller, a Neo M8N GPS receiver with a built-in compass, telemetry modules, Radio Transmitter-Receiver modules, and other related components. The Attacker Unit consists of a Raspberry Pi-4 model B to generate the spoofed GPS baseband signal data stream. A HackRF One Software Defined Radio (SDR) converts this data stream to Radio Frequency and transmits the spoofed signal with an antenna. The hardware architecture of the AVT is presented in Fig. \ref{avt}.

\subsubsection{Practical GPS Spoofing Attack on the AVT}

To perform the GPS spoofing attack on the AVT, a real-life spoofing attack scenario was generated based on the \textit{Attacker Model} outlined in Section \ref{attacker model} of this paper. The attack experiment utilized a single spoofed signal, mathematically denoted as $N_{spf}$ = 1. Consistent with the standard L1 GPS signal, the spoofed signal employed a carrier wave frequency of 1575.42 MHz. The navigation message contained information specific to a spoofed location, which was modulated onto the carrier wave using the BPSK modulation technique. The GPS satellite constellation was specified via a GPS broadcast ephemeris file (a collection of the individual site navigation files). The sampling frequency was set to 2600 MHz and a Gaussian noise of 20 MHz centered around 1575.42 MHz was generated for a stealthy attack. The spoofed signal was generated using the GPS-SDR-SIM tool \cite{gps-sdr-sim} and transmitted at a rate of 1 Hz.

To ensure ethical compliance and prevent interference with other GPS receivers or critical navigation systems, the transmission range and strength of the spoofed signal were strictly controlled and contained --- limiting it to less than 3 meters by controlling the transmit power. A test receiver was placed at various distances to measure the range. The transmit gain was set very low ($\approx$ -10 dB) and then gradually incremented while monitoring the spoofed signal's range. The gain setting at which the signal was detectable at 3 meters and dropped off significantly beyond that range was noted ($\approx$ 0 dB) and utilized in the experiments. The spoofing experiments were conducted in open, wide fields with no other vehicles or critical systems within range. As outlined in Algorithm \ref{spoofing algorithm}, the spoofing attack affected the AVT control system and compromised its normal state.

\subsubsection{Real Data Collection}

The real-world datasets were collected in two distinct locations separated by approximately 4 miles: The University of Arizona campus, and the Alvernon Park— both located in Tucson, Arizona, USA. %The field experiments were conducted in open and wide fields where no obstacle could hinder the autonomous operation of the vehicle. 
% A wireless remote controller was employed for manual control of the vehicle to avoid hazards and accidents. Nonetheless, the vehicle moved autonomously based on GPS guidance. 
During data collection, the vehicle was consistently operated in autonomous mode guided by GPS. As a safety precaution, a wireless remote controller was employed for manual control of the vehicle to mitigate potential hazards.

To collect normal data, a circular path consisting of 7 waypoints including the home location was mapped and uploaded to the AVT from the Ground Control Station. Each round of normal data collection was considered completed if the vehicle followed all 7 waypoints and returned to its home location. This way, 180 rounds of normal data were collected while no attack was imposed. To collect attack data, the AVT was operated in autonomous mode for 10 seconds without imposing any attack; afterward, the attack was launched. During one round of attack data collection, the spoofing attack was performed on the vehicle for 300 seconds. This way, 65 rounds of attack data were collected. By extracting the dataflash logs after each autonomous operation, all vehicular data were stored, labeled, and added to the dataset. To distinguish between the two categories of data, the normal and attack data were labeled by 0’s and 1’s, respectively.

\subsection{Simulation Experiments}

\subsubsection{CARLA Simulator} CARLA is an open-source urban driving simulator designed to facilitate the development, training, and validation of autonomous driving systems. CARLA provides a highly detailed, realistic environment that mimics urban settings, complete with dynamic weather conditions, diverse vehicle and pedestrian traffic, and a wide array of sensory inputs. %We used CARLA to run simulations in an urban environment and collect different vehicle sensor data including GNSS, accelerometer, and gyroscope. We used CARLA Town01 for the simulation. Town01 is a small town with some streets, few T-junctions, and a variety of buildings, surrounded by coniferous trees and featuring several small bridges spanning across a river that divides the town into two halves.
We utilized CARLA to run simulations in an urban environment where various vehicle sensor data was collected, including GPS signals, accelerometer readings, and gyroscope data.

\subsubsection{Simulation Environment}

The simulations were conducted in CARLA Town01, which is a small, urban-like area featuring a mix of streets, a few T-junctions, and a variety of buildings. The town is surrounded by coniferous trees and contains several small bridges that span a river, dividing the town into two halves. We ran 3 simulations, each lasting 5 minutes, to capture various randomizations of vehicle start locations. The simulations were performed in an urban town setting with clear weather conditions. In each simulation, 100 vehicles were spawned at different locations. Among these, 10 vehicles were equipped with GPS receivers, accelerometers, and gyroscopes. All vehicles in the simulations were controlled by an autopilot system to generate subsequent locations. The sensor data from the 10 equipped vehicles were monitored and logged at 0.1-second intervals.

\subsubsection{GPS Spoofing Attack Simulation} \label{spoofing_sim}
To simulate the GPS spoofing attack, we modified the GPS spoofed entries according to the equation: $ X' \sim \mathcal{N}(X + \mu,\,\sigma^{2})\,.$ where $X'$  is the GPS reading (latitude and longitude) after adding the spoofing effect, $X$ is the GPS reading before adding the spoofing effect, $\mu$ the mean of the GPS readings for each vehicle and $\sigma$ is the standard deviation of the GPS readings for each vehicle. This GPS spoofing method was used in \cite{shabbir2023securing}. We generated a GPS spoofed entry for each original entry.

\subsection{Autonomous Vehicle GPS Datasets (AV-GPS-Datasets)}\label{AV-GPS-Dataset}
%\textcolor{red}{The AV-GPS-Dataset is now made up of two separate components: 1) the car data, 2) simulation data; and in total has 6 datasets. So please make sure the dataset labels read correctly. AV-GPS-Dataset 1, 2,3; and then AV-GPS-Dataset4,5,6. Also some of the writeup in this section and section 5.B.3 have to be modified to reflect this. Also make sure the experiment results in the table reflect the correct names too}
The AV-GPS-Datasets have two parts with a total of 6 subsets, namely \textit{AV-GPS-Dataset (Real Data)}, with subsets 1---3, and \textit{AV-GPS-Dataset (Simulation-generated Data)}, with subsets 4---6. Each subset contains two classes of data: Normal and Attack data, and they are extracted as Comma Separated Value (CSV) files. The AV-GPS-Datasets mainly consist of temporal features of AVs collected using the AVT and CARLA simulated environment. %The features and labels are consistent across all datasets, ensuring uniformity in the categorization of the data.

\subsubsection{AV-GPS-Dataset 1---3 (Real Data)}

%The real part of the AV-GPS-Dataset has 3 subsets with varied entries, namely AV-GPS-Dataset 1, AV-GPS-Dataset 2, and AV-GPS-Dataset 3. 

%Each subset contains two classes of data: Normal and Attack data, and they are extracted as Comma Separated Value (CSV) files from the Ground Control Station computer. The features and labels are consistent across all datasets, ensuring uniformity in the categorization of the data. %Table \ref{tab1} presents a summary of the dataset along with brief descriptions of the features.

AV-GPS-Dataset 1 is the largest subset containing 62,042 entries, out of which, 46,787 are labeled as normal vehicular data (approximately 75\%), and 15,255 are labeled as attack data (approximately 25\%). This set contains experimental data from two different locations at the University of Arizona Campus. Both locations were chosen so that the AVT could have a clear reception of the GPS signal and no obstacle could block the GPS reception. Here, we collected the normal data and attack data in separate sessions, and afterward, merged the two types of data to form the dataset. The dataset comprises three distinct GPS spoofing scenarios- \textbf{Scenario I:} GPS spoofing when the AVT is following a straight line, \textbf{Scenario II:} GPS spoofing when the AVT is making turns, and \textbf{Scenario III:} GPS spoofing when the AVT is stationary. 

AV-GPS-Dataset 2 is the second subset and contains 6,890 entries, with 5,184 labeled as normal (approximately 75\%) and 1,706 labeled as attack data (approximately 25\%). This dataset was collected from outside the University of Arizona Campus (Alvernon Park), incorporating a different location and driving environment. This location had more trees to obstruct GPS reception partially, making the autonomous operation more challenging for the AVT. This subset was also collected in separate sessions, with normal and attack data merged to form the dataset. It comprises two spoofing scenarios, \textbf{Scenario I} and \textbf{Scenario III}.

AV-GPS-Dataset 3 is the smallest subset collected from the eastern part of the University of Arizona Campus. It contains only 636 entries, with 241 labeled as normal (approximately 38\%) and 395 labeled as attack data (approximately 62\%). This subset was collected in a single session to capture the transition between the normal and attack state of the AVT. The complete AV-GPS-Dataset is publicly available at \cite{AV-GPS-Dataset}, with a comprehensive explanation of the dataset features.

\subsubsection{AV-GPS-Dataset 4---6 (Simulation-generated Data)}

To prepare AV-GPS-Dataset 4---6 from CARLA simulations, the vehicle sensor data was first collected in normal conditions without imposing GPS spoofing attacks. Data from the 14 features highlighted in the Autonomous Vehicle Behavior Model was collected. %We used CARLA simulator to collect the simulation-generated data. The sensor data collected were the normal data (no GPS spoofing attack) that included the same features highlighted in the Autonomous Vehicle Behavior Model. %The 3 simulations resulted in 3 datasets: AV-GPS-Dataset 4, Dataset 5, and Dataset 6. Each dataset has a different random start position for each vehicle. 
Three simulations were conducted, resulting in three distinct datasets: AV-GPS-Dataset 4 with 29,735 entries of normal data, AV-GPS-Dataset 5 with 22,120 entries of normal data, and AV-GPS-Dataset 6 with 21,315 entries of normal data. Each dataset features different random starting positions for each vehicle. % and contains five minutes of simulation data collected at 0.1-second intervals. 
Traffic density varied across the datasets, with AV-GPS-Dataset 4 having heavy traffic density, AV-GPS-Dataset 5 having moderate traffic density, and AV-GPS-Dataset 6 having sparse traffic density. We then introduced GPS spoofing attacks by adding corresponding spoofed data entries as explained in \ref{spoofing_sim} and reran the experiments to collect the attack data, effectively doubling the number of data points in each dataset. As a result, AV-GPS-Dataset 4 contains a total of 59,470 data points, AV-GPS-Dataset 5 contains 44,240 data points, and AV-GPS-Dataset 6 contains 42,630 data points.

%Each dataset contains 5 minutes of simulation data with a data point collected every 0.1 seconds. That means we have around 3000 data points in each simulation-generated dataset since sometimes some vehicle sensors do not update on time,  the exact number of data points in AV-GPS-Dataset 4 is 29735, AV-GPS-Dataset 5 is 22120 and AV-GPS-Dataset 6 is 21315. We added the GPS spoofing attack data as described in subsection \ref{spoofing_sim}. We added a GPS spoofed entry for each normal data entry which doubled the number of data point in each dataset. 

\subsection{Experimental Analysis}

In this section, we present the experiments that were used to analyze the AV-GPS-Datasets and evaluate the performance of the GPS-IDS approach.

\subsubsection{Experiment 1- State Estimation of the AVT using the Autonomous Vehicle Behavior Model} 

This experiment establishes a connection between the Autonomous Vehicle Behavior Model and the actual AVT. In this experiment, we estimated the states of the AVT using the dynamic bicycle model presented in section \ref{section dynamic bicycle model}. The estimated results are then compared with the actual vehicle states obtained from 46,787 instances of normal data from AV-GPS-Dataset 1. Since a small-scale testbed is used to represent the real vehicle, it is assumed that the distances of the front and the rear wheel axle from the center of mass are equal. It is also assumed that the cornering stiffness coefficients for the front and rear wheels are equal to 1. Table \ref{physical parameters table} presents the physical parameters of the AVT that were used to identify the system matrix ``\textbf{a}" and the input matrix ``\textbf{b}". Substituting these values in Equation \ref{eq17}, the final dynamic state equation obtained is as follows:

\begin{equation}\label{final dynamics}
\begin{bmatrix}
\dot{v_y} \\
\dot{r}
\end{bmatrix} 
= 
\begin{bmatrix}
0.8 & 0.0 \\
0.0 & 1.1169
\end{bmatrix}
\begin{bmatrix}
v_y \\
r
\end{bmatrix}
+
\begin{bmatrix}
-0.4 \\
-2.5384
\end{bmatrix}
\delta
\end{equation}

\begin{figure}[b]
     \centering
     \begin{subfigure}{0.235\textwidth}
         \centering
         \includegraphics[width = \textwidth]{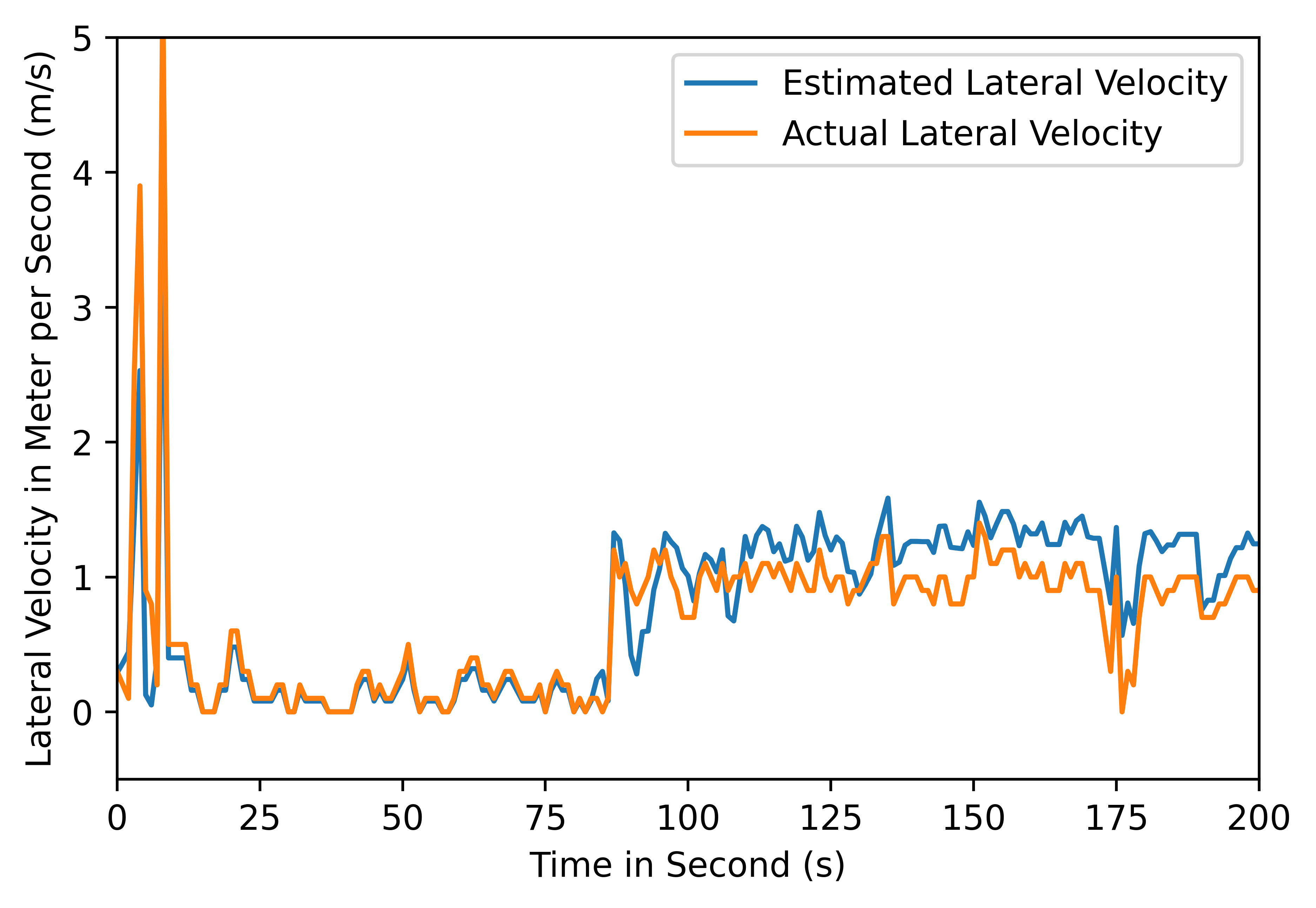}
         \caption{Estimated Lateral Velocity vs Actual Velocity}
         \label{estimated_lateral_velocity}
     \end{subfigure}
     \hfill
     \begin{subfigure}{0.245\textwidth}
         \centering
         \includegraphics[width = \textwidth]{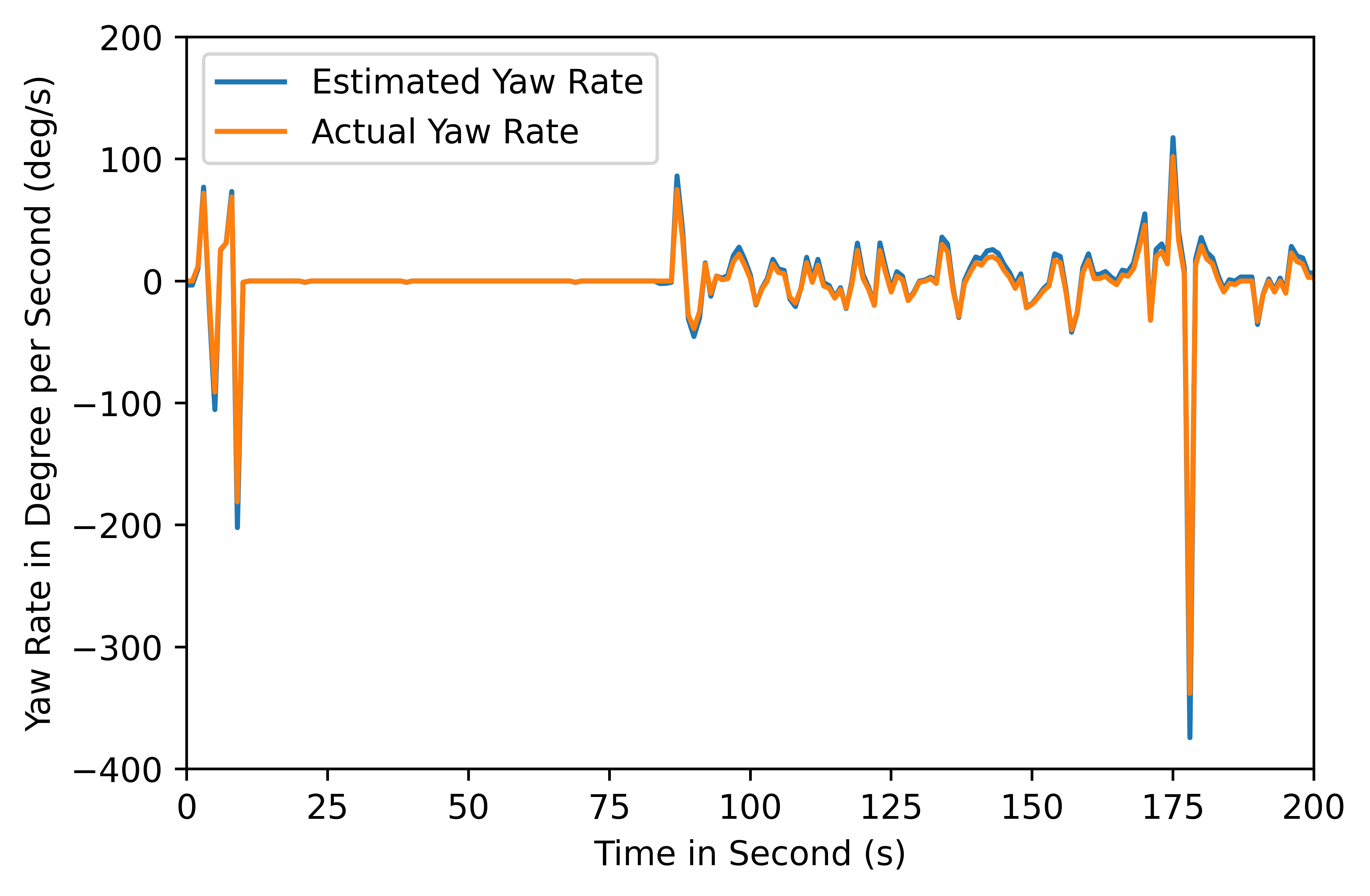}
         \caption{Estimated Yaw Rate vs Actual Yaw Rate}
         \label{Estimated Yaw Rate}
     \end{subfigure}
     \caption{Estimated Dynamics vs Actual Dynamics}
     \label{Estimation}
\end{figure}

\begin{table}[t]
\caption{Physical Parameters of the Autonomous Vehicle Testbed (AVT)}\label{physical parameters table}
\centering
\begin{tabular} {m{4cm} | m{2cm}}
\hline
\hline
\textbf{AVT Parameters} & \textbf{Values}\\
\hline
Mass, $m$ & 2.5 $kg$ \\
Length, $l$ & 0.56 $m$ \\
Width, $w$ & 0.32 $m$ \\
Yaw moment of inertia, $I_z$ & 0.0867 $kg-m^2$ \\
Distance between the front wheel axle and Center of Mass, $l_f$ & 0.22 $m$ \\
Distance between the rear wheel axle and Center of Mass, $l_r$ & 0.22 $m$ \\
Nominal velocity, $v$ & 1 $m/s$ \\
\hline
\hline
\end{tabular}
\end{table}

The estimated results of the lateral dynamics of the AVT compared with the actual values are shown in Fig. \ref{Estimation}. From Fig. \ref{Estimation}, it can be observed that the estimated states using the dynamic bicycle model follow a similar distribution as the actual values of the yaw rates and lateral velocities. Therefore, the derived dynamic bicycle model can represent the lateral dynamics of the AVT well. This points out the necessity of adapting the dynamic bicycle model to capture the behavior of AVs and collect data on the parameters identified by the vehicle behavior model. Since the presented vehicle model is capable of estimating the distributions of the next states of the AVT correctly, the state space representation accurately models the testbed. The differences in these values can be accepted based on the assumption that the field experiments added some noise that is not considered in the state equation.

%\begin{figure*}
%     \centering
%     \begin{subfigure}[b]{0.24\textwidth}
%         \centering
%         \includegraphics[width=\textwidth]{Position_Measurement_without_Imposing_Attack.png}
%         \caption{Position measurement}
%         \label{pose (a)}
%     \end{subfigure}
%     \hfill
%     \begin{subfigure}[b]{0.24\textwidth}
%         \centering
%         \includegraphics[width=\textwidth]{Orientation_Measurement_without_Imposing_Attack.png}
%         \caption{Orientation measurement}
%         \label{pose (b)}
%     \end{subfigure}
%     \hfill
%     \begin{subfigure}[b]{0.24\textwidth}
%         \centering
%         \includegraphics[width=\textwidth]{Cross_track_Error.png}
%         \caption{Controller input}
%         \label{controller (c)}
%     \end{subfigure}
%     \hfill
%     \begin{subfigure}[b]{0.24\textwidth}
%         \centering
%         \includegraphics[width=\textwidth]{Ground_Speed_without_Attack.png}
%         \caption{Controlled speed}
%         \label{controller (d)}
%     \end{subfigure}
%     \caption{AVT pose measurements and controller input/ output without imposing GPS spoofing attack}
%     \label{pose controller normal}
     %\end{figure}
%\vspace{0.5cm}

\begin{figure*}
    \centering
     \begin{subfigure}[b]{0.24\textwidth}
         \centering
         \includegraphics[width=\textwidth]{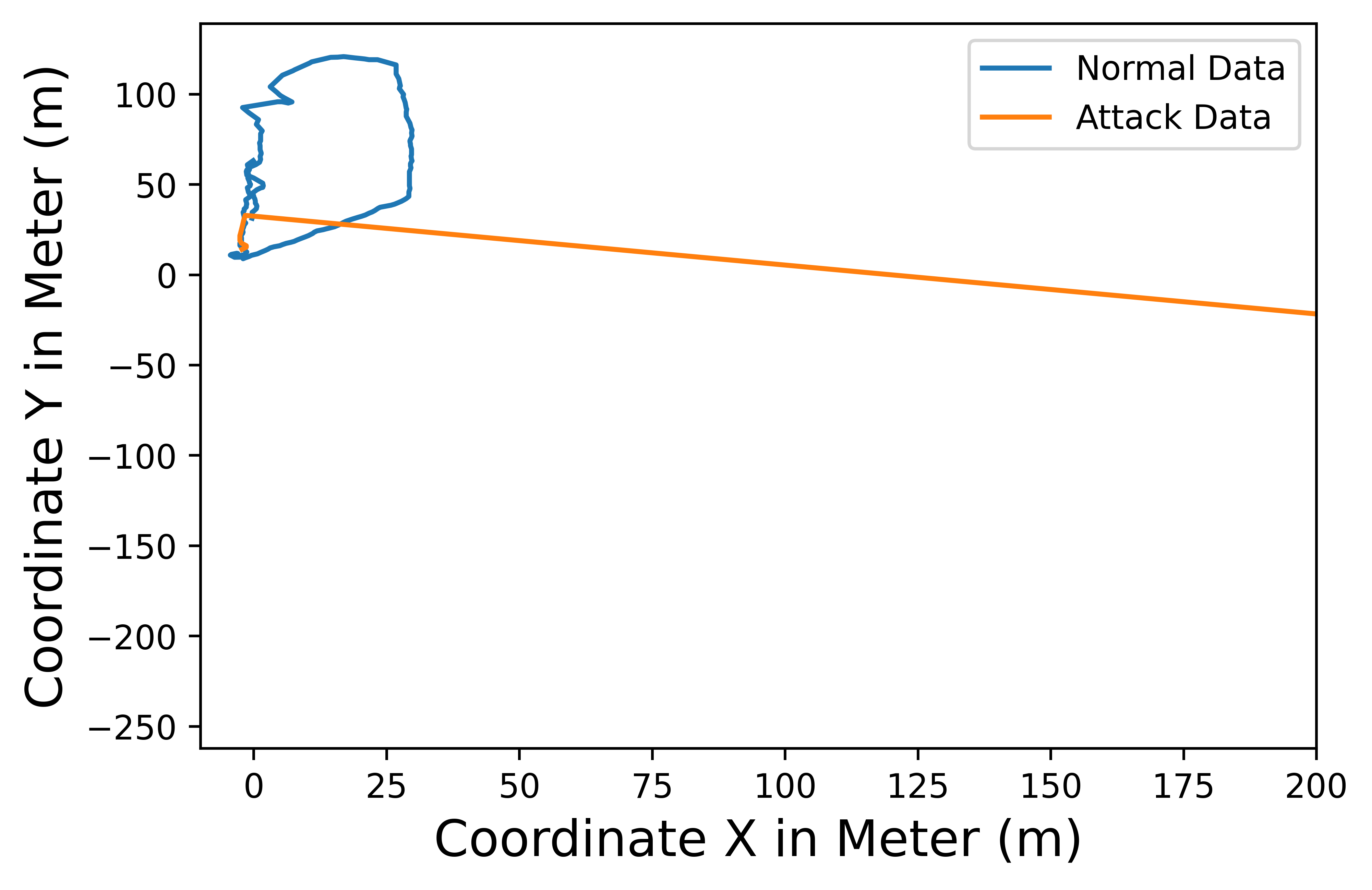}%Position_Measurement_during_Attack.png}
         \caption{Position measurement}
         \label{attack pose (a)}
     \end{subfigure}
     \hfill
     \begin{subfigure}[b]{0.24\textwidth}
         \centering
         \includegraphics[width=\textwidth]{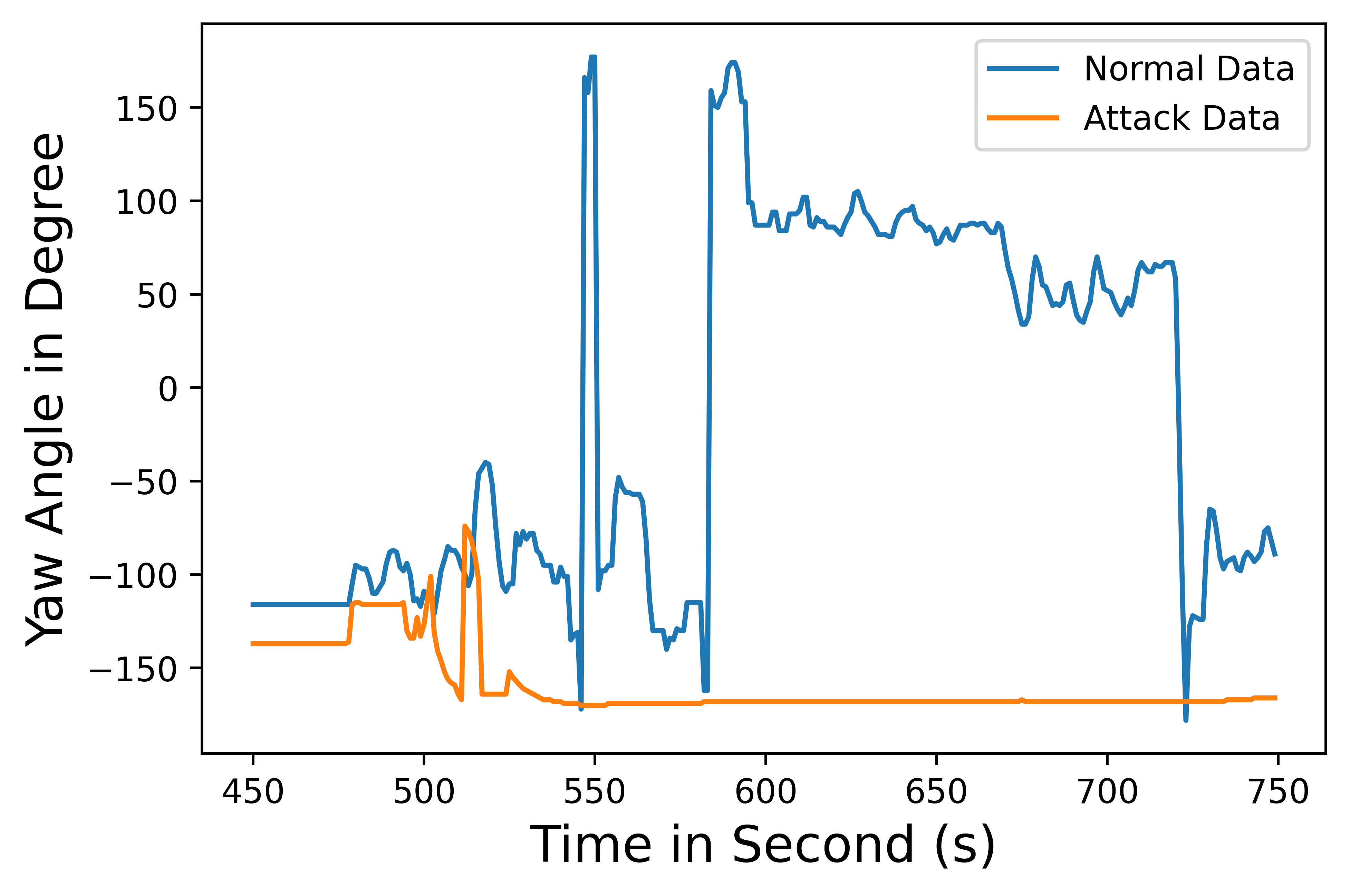}
         \caption{Orientation measurement}
         \label{attack pose (b)}
     \end{subfigure}
     \hfill
     \begin{subfigure}[b]{0.24\textwidth}
         \centering
         \includegraphics[width=\textwidth]{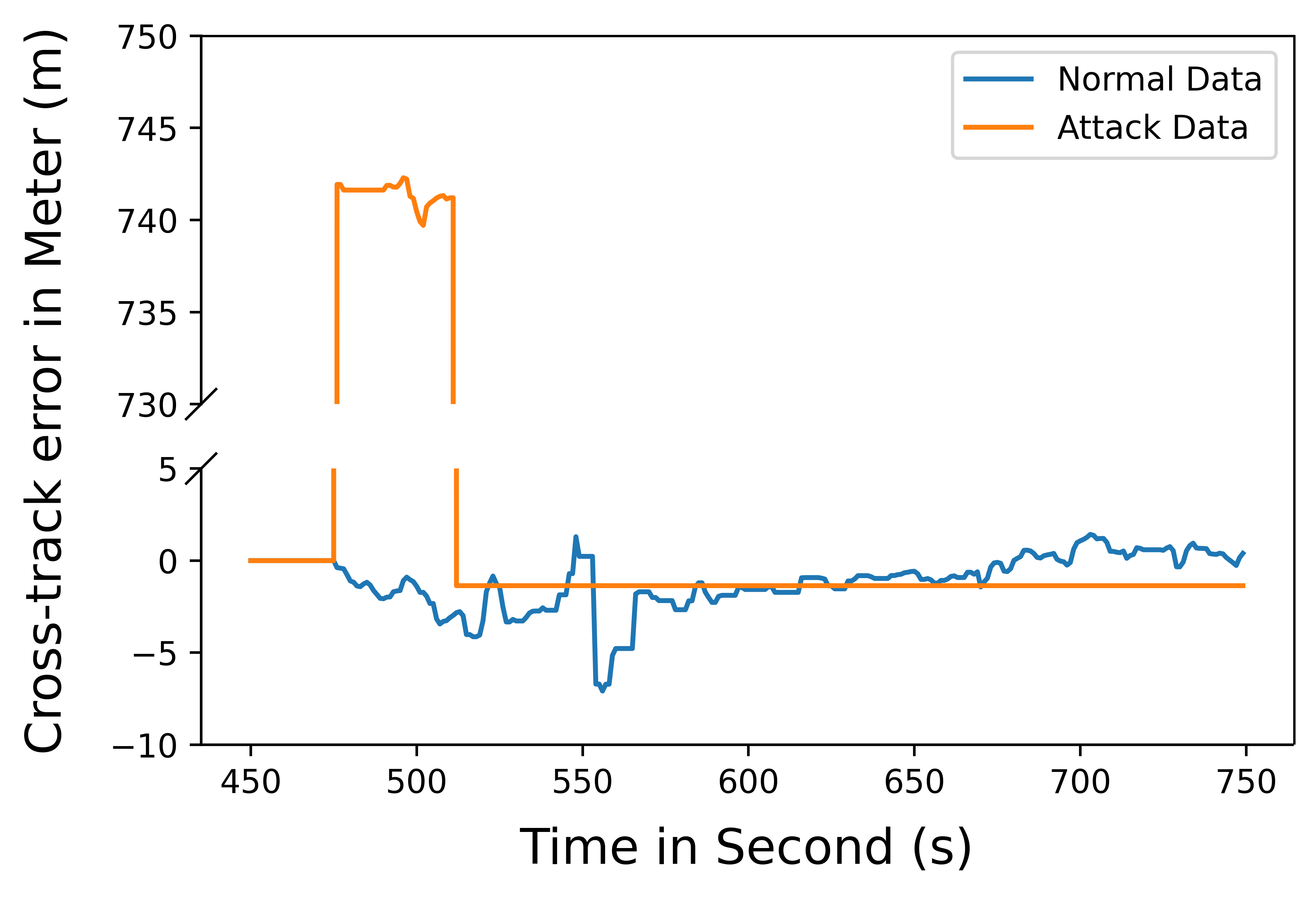}
         \caption{Controller input}
         \label{attack controller (c)}
     \end{subfigure}
     \hfill
     \begin{subfigure}[b]{0.24\textwidth}
         \centering
         \includegraphics[width=\textwidth]{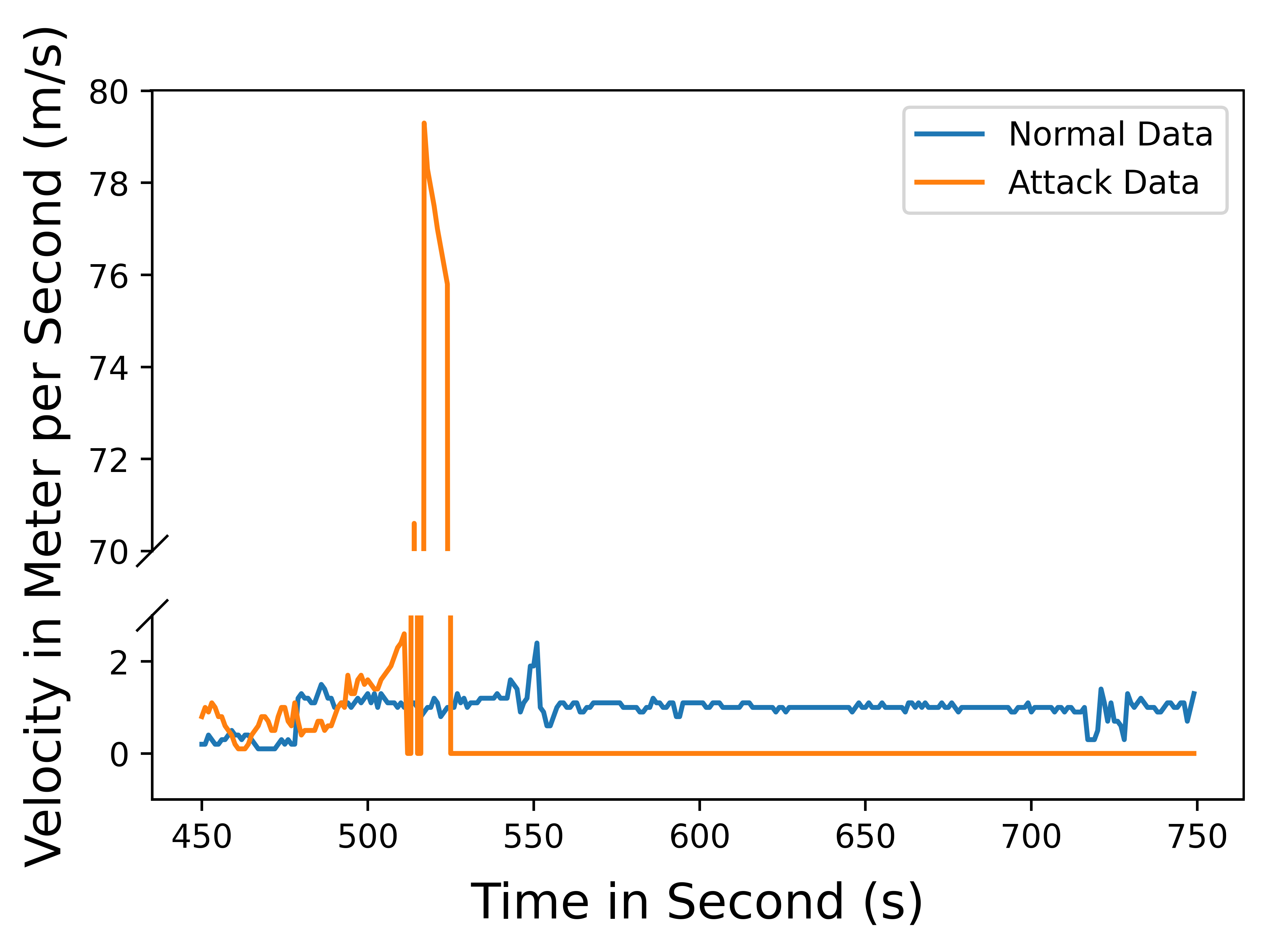}
         \caption{Controlled speed}
         \label{attack controller (d)}
     \end{subfigure}
     \caption{AVT pose measurements and controller input/ output under the influence of GPS spoofing attack. Axis breaks are utilized to shrink down large segments and enhance readability}
     \label{pose controller attack}
\end{figure*}

\begin{figure*}%[htbp]
\centerline{\includegraphics[width=17cm]{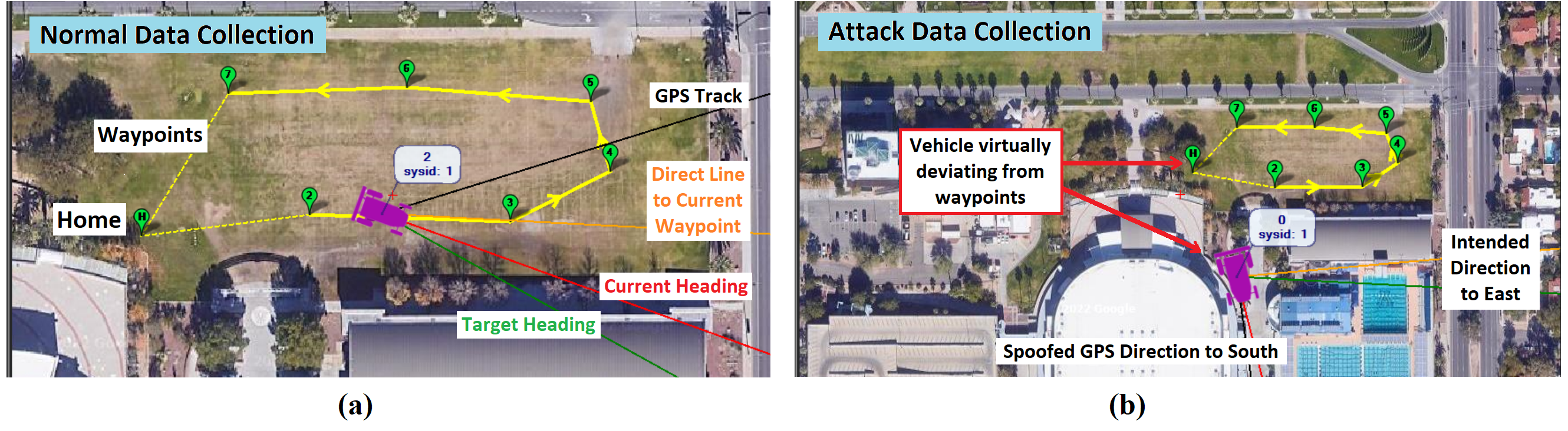}}
\caption{Screenshot of the data collection procedure from the Ground Control Station computer. Fig. (a) AVT following given waypoints during normal data collection, Fig. (b) AVT virtually moving to spoofed GPS location during attack data collection}\label{gcs screenshot}
\end{figure*}

% We can observe that the AVT follows a circular trajectory with a velocity ranging from 0 m/s to 2.5 m/s. During its path traversal, it effectively corrects any cross-track errors and adjusts its yaw angles accordingly. 

\subsubsection{Experiment 2- Attack Impact on Pose Measurements and Controller Input/ Output} 

This experiment presents a comprehensive analysis of the behavior of the AVT under normal operating conditions and under the influence of real GPS spoofing attack. To facilitate a clear comparison of the attack impact, we plotted one round of normal data with the corresponding attack data from AV-GPS-Dataset 1 pertaining to the pose measurements and controller input/ output of the AVT, as presented in Fig. \ref{pose controller attack}. %We initially depicted the AVT's normal behavior in terms of pose measurements and controller input/ output by plotting one round of normal data, as shown in Fig. \ref{pose controller attack}. %Fig. \ref{pose (a)} and Fig. \ref{pose (b)} illustrate the AVT's position and orientation, while Fig. \ref{controller (c)} and Fig. \ref{controller (d)} illustrate the cross-track error and velocity, respectively, under normal operating conditions. From Fig. \ref{pose controller normal},
We can observe that the AVT follows a circular trajectory (\ref{attack pose (a)}) with a velocity ranging from 0 m/s to 2.5 m/s. During its path traversal, it effectively corrects cross-track errors and adjusts its yaw angles accordingly (\ref{attack controller (d)}).% To facilitate a clear comparison of the attack impact, we plotted one round of normal data with the corresponding attack data pertaining to the pose measurements and controller input/ output of the AVT. %as presented in Fig. \ref{pose controller attack}.

It is apparent that the AVT's position is significantly affected by the spoofed GPS signals. It deviates from its regular circular trajectory and moves toward a different direction along the X coordinate (Fig. \ref{attack pose (a)}) at an irregular velocity spiking to as high as 80 m/s, and then decelerating to 0 m/s (Fig. \ref{attack controller (d)}). In Fig. \ref{attack controller (c)}, we can observe an abnormal rise in cross-track errors exceeding 740 meters, then dropping to around 0 meters after 35 seconds and continuing to be the same. The AVT was unable to correct this error and adjust its heading, causing the yaw angle to remain unchanged (Fig. \ref{attack pose (b)}). During this time, the AVT had lost all the satellite locks and remained stationary due to the variance in EKF estimates. Nonetheless, on the Ground Control Station map, the AVT was observed virtually progressing toward the location indicated by the spoofed signals, which is shown in Fig. \ref{gcs screenshot}.

%%%%%%%%%%%%%_BEGIN_TABLE_%%%%%%%%%%%%%%%%

\newcommand{\STAB}[1]{\begin{tabular}{@{}c@{}}#1\end{tabular}}

\begin{table*}
\caption{Performance of the Machine Learning Classification Models on AV-GPS-Datasets 1---3 (Real Data)}\label{tab2}
  \centering
  \begin{tabular}{|c|c|c|c|c|c|c|c|c|c|c|} 
    \hline
    & \textbf{AV-GPS-Datasets 1---3} & \textbf{Metrics} & \textbf{RF} & \textbf{XGB} & \textbf{SVC} & \textbf{MLP} & \textbf{Adaboost} & \textbf{GB} & \textbf{DT} \\
    \hline
    \multirow{10}{*}{\STAB{\rotatebox[origin=c]{90}{\textbf{Case I}}}}
    \multirow{10}{*}{\STAB{\rotatebox[origin=c]{90}{Trained with 80\% of Dataset 1}}}
    & \multirow{4}{*}{Test on 20\% of Dataset 1} 
    
    & Accuracy & 0.972 & 0.965 & 0.972 & 0.974 & 0.867 & 0.970 &    0.933  \\ 

    & & Precision & 0.994 & 0.969 & 0.985 & 0.979 & 0.671 & 0.962 &   0.810    \\
    
    & & Recall & 0.899 & 0.893 & 0.903 & 0.919 & 0.935 & 0.919 &   0.965   \\ 
    
    & & F1 Score & 0.944 & 0.929 & 0.942 & 0.948 & 0.782 & 0.939 &    0.881   \\ 
    
    \cline{2-10}

    & \multirow{4}{*}{Test on Dataset 2} & Accuracy & 0.975 & 0.979 & 0.999 & 0.995 & 0.250 & 0.842 & 0.861    
    \\ 
    
    & & Precision & 0.912 & 0.933 & 0.999 & 0.984 & 0.248 & 0.610 &  0.640    \\ 
    
    & & Recall & 0.997 & 0.986 & 0.997 & 0.996 & 1 & 1 &  0.997    \\ 
    
    & & F1 Score & 0.953 & 0.959 & 0.998 & 0.990 & 0.397 & 0.758 &  0.780 \\ 
    \cline{2-10}

    & \multirow{4}{*}{Test on Dataset 3} & Accuracy & 0.965 & 0.945 & 0.963 & 0.948 & 0.636 & 0.874 & 0.902 \\ 

    & & Precision & 0.997 & 1 & 0.997 & 0.967 & 0.636 & 0.839 & 0.883  
   \\ 
    
    & & Recall & 0.948 & 0.913 & 0.945 & 0.950 & 1 & 0.992 & 0.975    \\ 
    
    & & F1 Score & 0.972 & 0.954 & 0.970 & 0.958 & 0.778 & 0.909 &   0.927    \\

    \hline

    \multicolumn{3}{|c|}{Average F1 Score in Case I} & 0.956   &  0.947 & \textbf{0.970} & 0.965 & 0.652 & 0.869 &   0.862     \\

    \hline

    %%%%%%%%%%%%%%%%%%%%%%%%%%%%%%%%%%%%%%%%%%%%%%%%%%%%%%%%%%%
    \multirow{10}{*}{\STAB{\rotatebox[origin=c]{90}{\textbf{Case II}}}}
    \multirow{10}{*}{\STAB{\rotatebox[origin=c]{90}{Trained with 80\% of Dataset 2}}}
    & \multirow{4}{*}{Test on Dataset 1} & Accuracy & 0.891 & 0.890 & 0.760 & 0.964 & 0.891 & 0.982 &   0.973   \\
    
    & & Precision & 0.992 & 0.994 & 0.980 & 0.996 & 0.992 & 0.978 &   0.997    \\
    
    & & Recall & 0.576 & 0.570 & 0.058 & 0.861 & 0.577 & 0.952 &    0.899    \\
    
    & & F1 Score & 0.729 & 0.725 & 0.109 & 0.924 & 0.730 & 0.965 &    0.946   \\
    \cline{2-10}

    & \multirow{4}{*}{Test on 20\% of Dataset 2} & Accuracy & 0.998 & 0.998 & 0.998 & 0.996 & 0.974 & 0.970 &  0.922       \\ 

    & & Precision & 0.998 & 0.998 & 0.994 & 0.998 & 0.998 & 0.961 &    0.926   \\
    
    & & Recall & 0.996 & 0.996 & 0.995 & 0.985 & 0.897 & 0.919 &   0.753    \\ 
    
    & & F1 Score & 0.997 & 0.997 & 0.997 & 0.992 & 0.945 & 0.939 &     0.830   \\ 
    \cline{2-10}

    & \multirow{4}{*}{Test on Dataset 3} & Accuracy & 0.973 & 0.971 & 0.963 & 0.971 & 0.973 & 0.874 &    0.965    \\ 
    
    & & Precision & 1 & 1 & 0.997 & 0.997 & 0.997 & 0.839 & 1      \\ 
    
    & & Recall & 0.958 & 0.955 & 0.945 & 0.958 & 0.960 & 0.992 &     0.945    \\ 
    
    & & F1 Score & 0.978 & 0.977 & 0.970 & 0.977 & 0.978 & 0.909 &  0.972     \\
    \hline

    \multicolumn{3}{|c|}{Average F1 Score in Case II} & 0.902 & 0.900 & 0.692 & \textbf{0.964} & 0.884 & 0.938 &    0.916   \\
    
    \hline

    %%%%%%%%%%%%%%%%%%%%%%%%%%%%%%%%%%%%%%%%%%%%%%%%%
    
    \multirow{10}{*}{\STAB{\rotatebox[origin=c]{90}{\textbf{Case III}}}}
    \multirow{10}{*}{\STAB{\rotatebox[origin=c]{90}{Trained with 80\% of Dataset 3}}}
    & \multirow{4}{*}{Test on Dataset 1} & Accuracy & 0.891 & 0.891 & 0.760 &    0.903   & 0.739 & 0.766 &   0.888    \\ 
    
    & & Precision & 0.983 & 0.982 & 0.939 &  0.904    & 0.487 & 0.536 &    0.982    \\ 
    
    & & Recall & 0.581 & 0.582 & 0.061 &    0.693   & 0.546 & 0.586 &    0.570   \\ 
    
    & & F1 Score & 0.730 & 0.731 & 0.115 &   0.784   & 0.515 & 0.560 &   0.721    \\ 
    \cline{2-10}

    & \multirow{4}{*}{Test on Dataset 2} & Accuracy & 0.998 & 0.998 & 0.807 &   0.990    & 0.953 & 0.973 &    0.995   \\ 
    
    & & Precision & 0.995 & 0.993 & 0.979 &  0.961   & 0.842 & 0.903 &     0.985   \\ 
    
    & & Recall & 1 & 1 & 0.228 &   1   & 1 & 1 &  0.998    \\ 
    
    & & F1 Score & 0.997 & 0.996 & 0.369 &   0.980   & 0.914 & 0.949 &   0.991     \\ 
    \cline{2-10}

    & \multirow{4}{*}{Test on 20\% of Dataset 3} & Accuracy & 0.962 & 0.959 & 0.945 &   0.940   & 0.952 & 0.951 &  0.971     \\ 
    
    & & Precision & 0.984 & 0.965 & 0.967 &   0.946    & 0.963 & 0.953 &    1     \\ 
    
    & & Recall & 0.955 & 0.970 & 0.945 &   0.960    & 0.963 & 0.970 &      0.955    \\ 
    
    & & F1 Score & 0.969 & 0.968 & 0.956 &   0.953   & 0.963 & 0.962 &    0.977    \\
    \hline

    \multicolumn{3}{|c|}{Average F1 Score in Case III} & 0.899 & 0.898 & 0.480 &  \textbf{0.905}  & 0.797 & 0.823 &   0.896     \\

    \hline

    \multicolumn{3}{|c|}{\textbf{Average F1 Score of all three cases}} & 0.919 & 0.915 & 0.714 &  \textbf{0.944}  & 0.777 & 0.876 &   0.891    \\
    
    \hline

  \end{tabular}
\end{table*}

\subsubsection{Experiment 3- Performance Analysis of the Machine Learning Models with Different Training Sets from Real Data}\label{ABA_exp3}

This experiment evaluated the GPS-IDS approach on the AV-GPS-Dataset 1---3 using different machine learning models. To benchmark various machine learning techniques, including ensemble methods, neural networks, and tree-based algorithms, seven models were chosen for detection: Random Forest (RF), XGBoost (XGB), Support Vector Machine Classifier (SVC), Multi-Layer Perceptron (MLP), AdaBoost, Gradient Boosting (GB), and Decision Tree (DT). The 14 features associated with the physics-based component of the Autonomous Vehicle Behavior Model (Fig. \ref{Autonomous Vehicle Behavior Model}) are utilized in the machine learning analysis. %Features associated with equations \ref{eq17} (Vehicle Model block), \ref{eq18} (Localization block), \ref{eq19} (State Estimation block), \ref{eq20} (Motion Planning block), and \ref{eq22} (PID Controller block) from the Autonomous Vehicle Behavior Model (Fig. \ref{Autonomous Vehicle Behavior Model}) are extracted and utilized in the machine learning models. %More details about feature selection can be found at \cite{AV-GPS-Dataset}.
To maintain the proportion of samples for each class consistently across both training and testing phases, stratified 5-fold sampling was employed to partition the data into training and testing sets. The training approach is based on supervised learning, where 4 folds (80\%) were utilized for training, and the remaining fold (20\%) was utilized for testing. Three cases were considered to train the models, \textbf{Case I:} Train on 80\% of AV-GPS-Dataset 1, and test on 20\% of AV-GPS-Dataset 1, AV-GPS-Dataset 2 and AV-GPS-Dataset 3; \textbf{Case II:} Train on 80\% of AV-GPS-Dataset 2, and test on AV-GPS-Dataset 1, 20\% of AV-GPS-Dataset 2 and AV-GPS-Dataset 3; \textbf{Case III:} Train on 80\% of AV-GPS-Dataset 3, and test on AV-GPS-Dataset 1, AV-GPS-Dataset 2 and 20\% of AV-GPS-Dataset 3. For all three cases, the performance of the machine learning models is measured in terms of Accuracy, Precision, Recall, and F1-score. For each model, a combination of hyperparameters was utilized and fine-tuned to find the optimal performance. The performance of machine learning algorithms on different datasets is presented in Table \ref{tab2}. The results indicate that the performance varies across different datasets as the training set changes. It can be observed that Case I exhibits high accuracy and F1 scores for most models, while the F1 scores drop for most models in Case III. This drop can be attributed to the smaller size of the training set in Case III, which contains a limited number of entries representing only the transition of the attack. In this case, the test sets are significantly larger than the training set (AV-GPS-Dataset 1 is over 110 times larger, and AV-GPS-Dataset 2 is over 12 times larger), which declines the model performances. Overall, MLP consistently performs with F1 scores above 90\% in all three cases, achieving the highest average F1 score of 94.4\%. Following closely, RF and XGB demonstrated similar and second-best performances, achieving average F1 scores of 91.9\% and 91.5\%, respectively, in the three considered cases. The DT classifier secured the third-best performance, attaining an average F1 score of 89.1\% across the three cases.

\subsubsection{Experiment 4- Tuning of the Detection Threshold $\mathbb{T}$} \label{threshold_tuning}

\begin{figure}[t]
\centerline{\includegraphics[width=7.5cm]{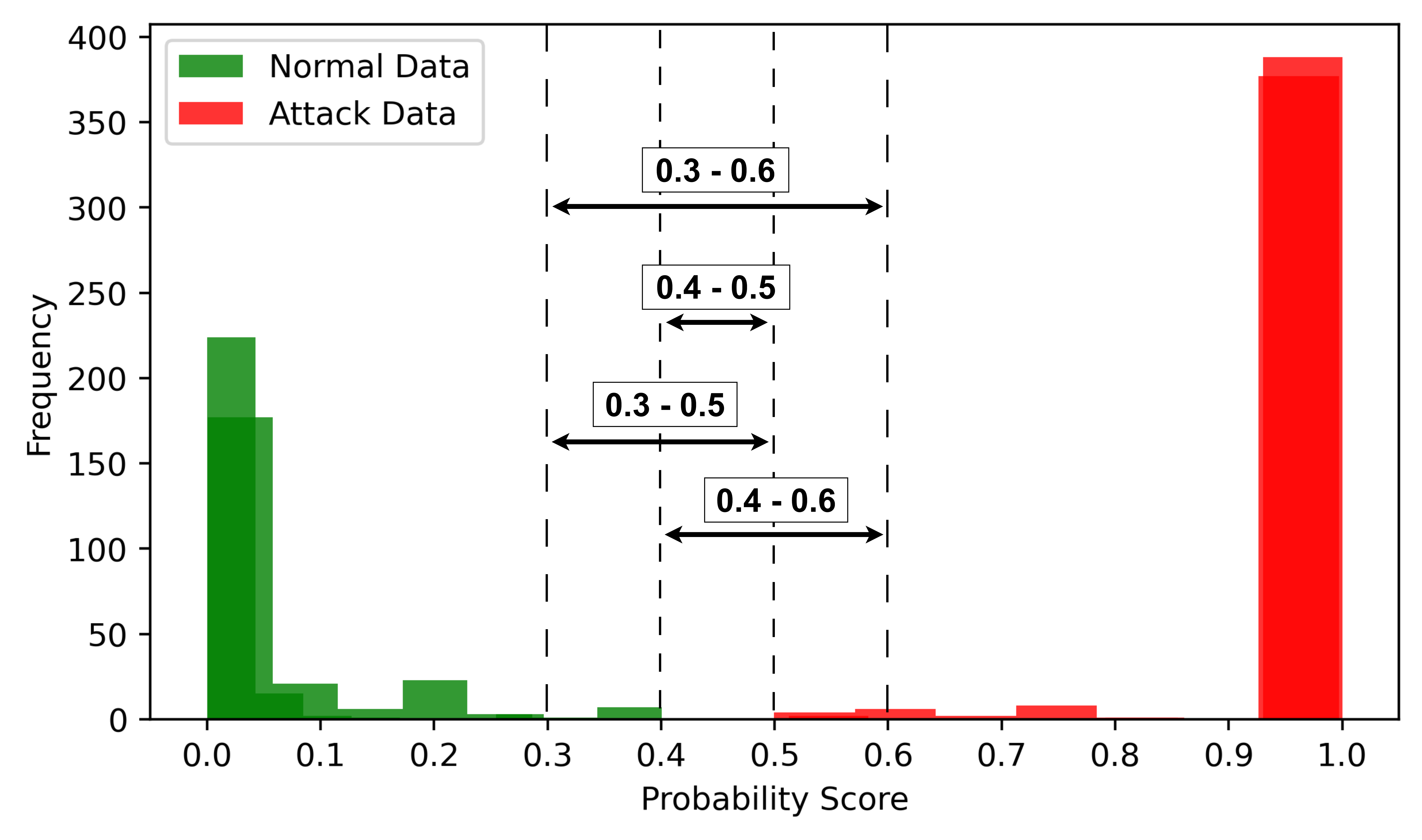}}
\caption{Comparison of probability score distribution for normal and attack data}\label{prob_score_margin}
\end{figure}

In this experiment, we conducted an analysis to optimize the predefined detection threshold $\mathbb{T}$ of the GPS-IDS framework by evaluating the probability scores of normal data and attack data from AV-GPS-Dataset 3. The machine learning model that performed the best in Experiment 3, namely MLP, was employed for this purpose. The resulting data were plotted to determine a suitable detection margin, as illustrated in Fig. \ref{prob_score_margin}. Fig. \ref{prob_score_margin} depicts the probability score distributions for normal and attack data obtained by applying Case I and Case II training methodologies on the same graph. By comparing the distributions, it becomes evident that the two kinds of data can be effectively distinguished with a significant margin indicated by the dotted bars. Table \ref{FPFN} shows the False Positive and False Negative Rates as we expand the margin. Based on the findings presented in Table \ref{FPFN}, we can see that selecting a detection threshold $\mathbb{T}$ within a range of \textbf{0.4 to 0.5} produces optimal outcomes. This range leads to the misclassification of only 1 instance of normal data and 1 instance of attack data, resulting in a False Positive Rate of 0.0021 or 0.21\% and a False Negative Rate of 0.0012 or 0.12\%. Thus, it can be deduced that opting for a detection threshold $\mathbb{T}$ in the range of 0.4 to 0.5 offers the lowest incidence of false detection and misclassification.

\begin{table}[t]
\caption{False Positive (FP) and False Negative (FN) Rates for Different Detection Margins in Real Datasets}\label{FPFN}
\centering
\begin{tabular}{c|c|c|c|c}
\hline
\hline
\multirow{2}{*}{\begin{tabular}[c]{@{}c@{}}Detection\\ Margin \end{tabular}} & \multirow{2}{*}{\begin{tabular}[c]{@{}c@{}}Normal Data\\ Misclassified\end{tabular}} & \multirow{2}{*}{\begin{tabular}[c]{@{}c@{}}Attack Data\\ Misclassified\end{tabular}} & \multirow{2}{*}{\begin{tabular}[c]{@{}c@{}}FP \\ Rate\end{tabular}} & \multirow{2}{*}{\begin{tabular}[c]{@{}c@{}}FN\\ Rate\end{tabular}} \\
 &  &  &  &  \\
%Detection Margin Range & Normal Data Misclassified & Attack Data Misclassified & False Positive Rate & False Negative Rate \\
\hline
\textbf{0.4 - 0.5} & \textbf{1} & \textbf{1} & \textbf{0.0021} & \textbf{0.0012} \\
0.3 - 0.5 & 7 & 1 & 0.0146 & 0.0012 \\
0.4 - 0.6 & 1 & 7 & 0.0021 & 0.0087 \\
0.3 - 0.6 & 7 & 7 & 0.0146 & 0.0087  \\
\hline
\hline
\end{tabular}
\end{table}

%Upon comparing the distributions depicted in the figure, it becomes apparent that the two categories of data can be effectively discerned with a notable distinction, as evidenced by the presence of the dotted bars. When the detection threshold (T) is set within the range of 0.4 to 0.5, 2 instances of normal data are misclassified, resulting in a false positive rate of 0.0041 or 0.41%. Expanding the range of T to 0.3 to 0.5 leads to misclassification of 9 normal data points, yielding an approximate false positive rate of 0.0183 or 1.83%. Setting T within 0.4 to 0.6 results in the misclassification of 2 normal data points and 4 attack data points, consequently leading to a false positive rate of approximately 0.0041 or 0.41% and a false negative rate of approximately 0.0051 or 0.51%. Setting T within 0.3 to 0.6 causes the misclassification of 9 normal data points and 4 attack data points, yielding an approximate false positive rate of 0.0183 or 1.83% and a false negative rate of approximately 0.0051 or 0.51%. Thus, it can be inferred that selecting a detection threshold T in the range of 0.4 to 0.5 affords the lowest occurrence of false detections.

\subsubsection{Experiment 5- Time Series Analysis of the Machine Learning Models}

\begin{figure}
     \centering
     \begin{subfigure}[b]{0.4\textwidth}
         \centering
         \includegraphics[width=\textwidth]{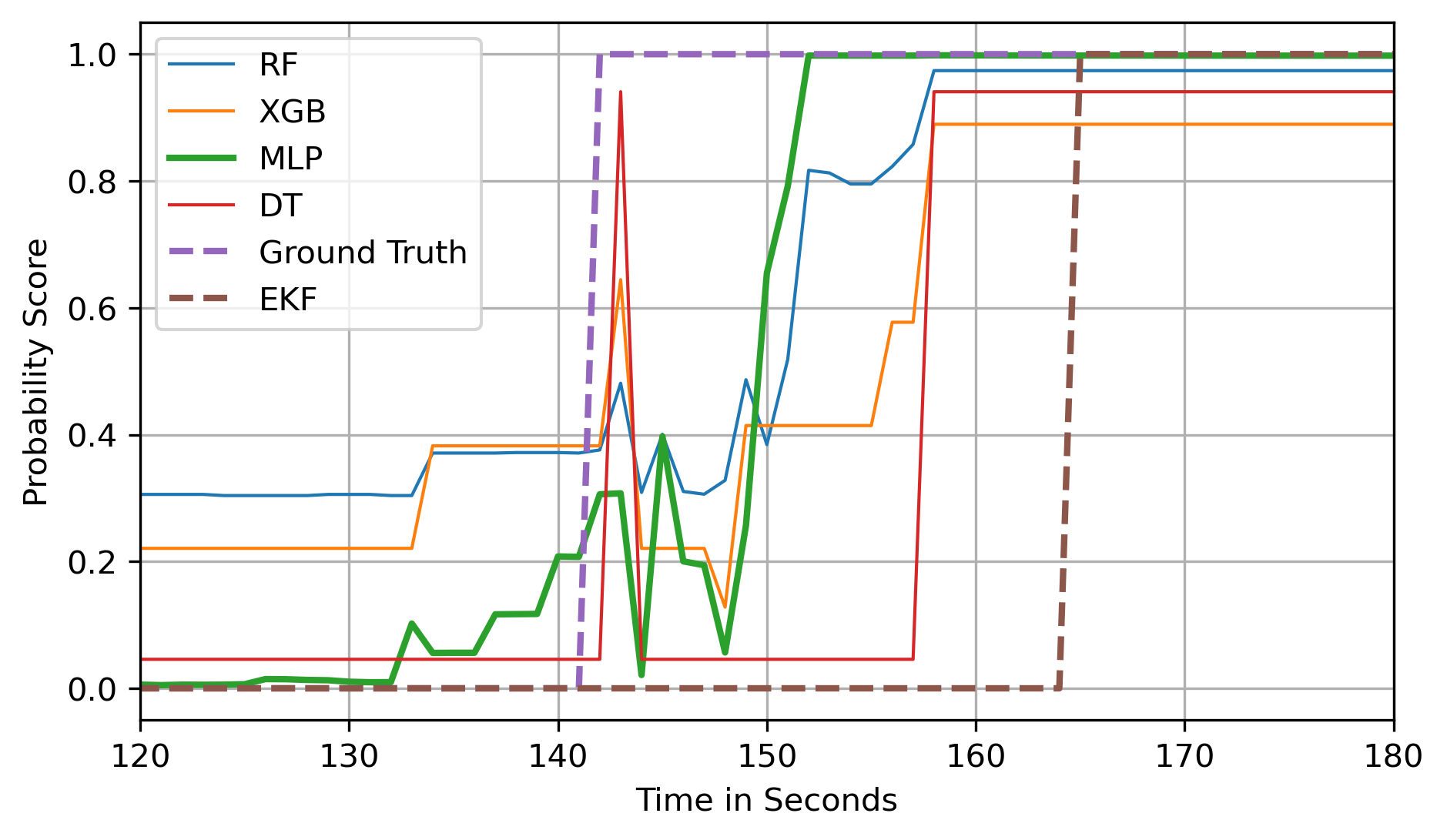}
         \caption{Using Case I training methodology}
         \label{fig11a}
     \end{subfigure}
     \hfill
     \begin{subfigure}[b]{0.4\textwidth}
         \centering
         \includegraphics[width=\textwidth]{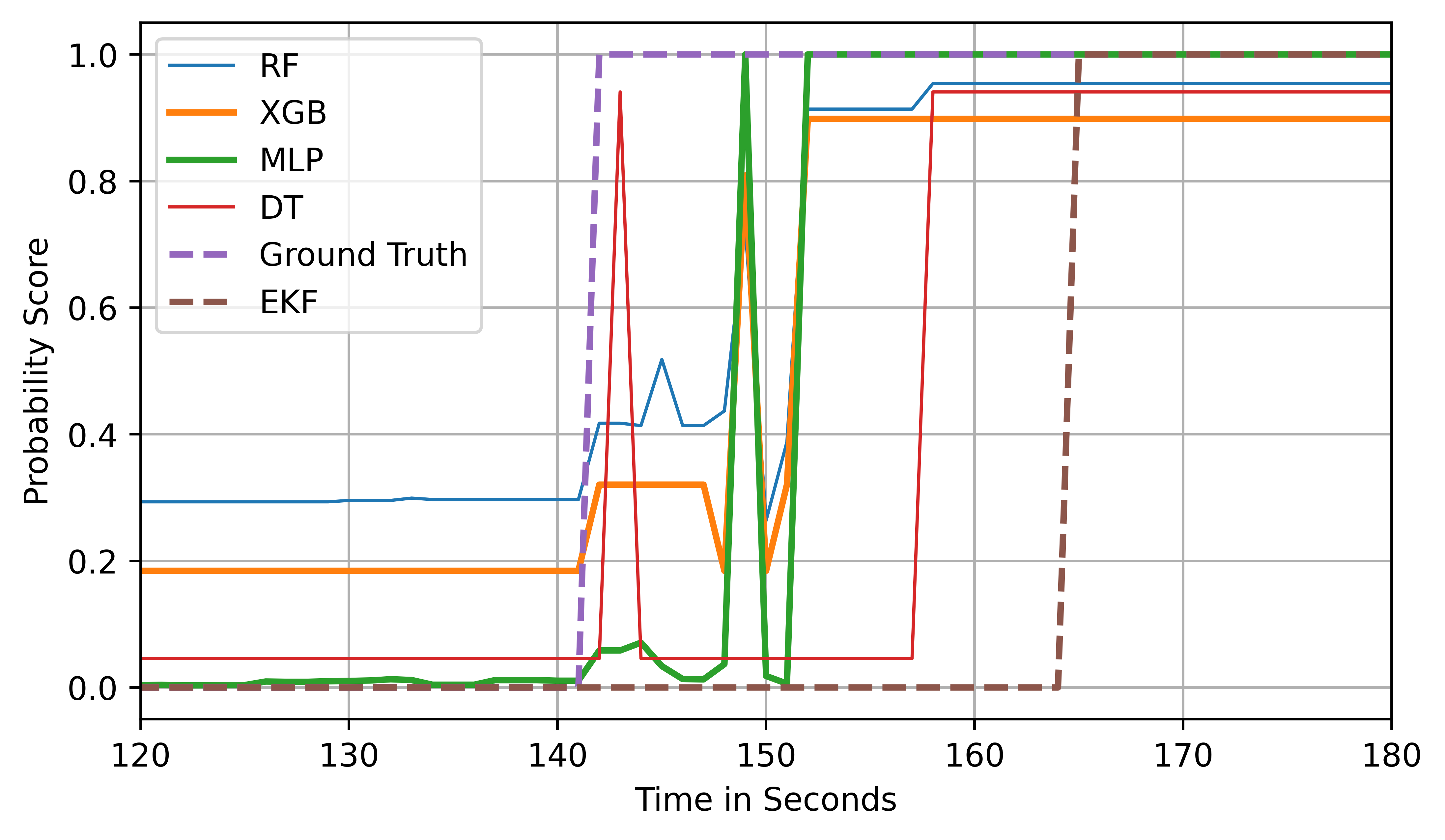}
         \caption{Using Case II training methodology}
         \label{fig11b}
     \end{subfigure}
     \caption{Time series analysis on AV-GPS-Dataset 3}\label{fig11}
\end{figure}

In this experiment, we evaluated the effectiveness of the 4 best-performing models from Experiment 3 (MLP, RF, XGB, and DT) in detecting an attack with respect to time on AV-GPS-Dataset 3. The training phase involved training the models using Case I and Case II methodologies, followed by testing on AV-GPS-Dataset 3. The outcomes of the time series analysis are represented in Fig. \ref{fig11}. As depicted in Fig. \ref{fig11}, the attack was initiated in the $142^{th}$ second and was subsequently detected by the AVT's EKF algorithm in the $165^{th}$ second (23 seconds delay). In both training cases, all the models succeeded in detecting the attack before the EKF. In Fig. \ref{fig11a}, MLP was capable of detecting the attack at the $152^{nd}$ second, followed by RF, XGB, and DT at the $158^{th}$ second. Similarly, in Fig. \ref{fig11b}, the MLP, XGB, and RF identified the attack at the $152^{nd}$ second, while the DT detected it at the $158^{th}$ second. Table \ref{tab5} summarizes the attack detection times for each classifier, revealing that MLP achieved the fastest detection time of 10 seconds in both training methodologies and showcasing a $\approx$ 56.5\% improvement in detection time compared to GPS/INS-based detection method using EKF. 

While a 10-second detection delay is considerably high in AV applications, it is important to highlight that the attack requires approximately 4-7 seconds to jam the reception of the authentic GPS signal and be effective on the AV. There are scopes for further improving the detection delay by fusing multiple detection techniques and increasing granularity. 

\begin{table}[t]
\caption{Results of the Time Series Analysis}\label{tab5}
\centering
\begin{tabular}{c|c|c|c|c|c|c}
\hline
\hline
\multicolumn{6}{c}{\hspace{2cm} Attack Detection Time (in seconds)} \\
\hline
 & \textbf{MLP} & \textbf{XGB} & \textbf{RF} & \textbf{DT} & \textbf{EKF} & \textbf{Max. Improve}\\
\hline
\textbf{Case I} & 10 & 16 & 16 & 16 & 23 & $\approx$ 56.52\%\\
\textbf{Case II} & 10 & 10 & 10 & 16 & 23 & $\approx$ 56.52\%\\
\hline
\hline
\end{tabular}
\end{table}

\subsubsection{Experiment 6- Performance Analysis of the Machine Learning Models with Different Simulated Training Sets} \label{sim_ML}

To ensure the generalization of the GPS-IDS approach, this experiment evaluated GPS-IDS on the simulation-generated AV-GPS-Datasets 4---6 using different machine learning models. Six models were chosen: RF, XGB, SVC, MLP, AdaBoost, and GB. Similar to Experiment 3, supervised learning was utilized and stratified 5-fold sampling was employed, where 4 folds (80\%) were used for training, and the remaining fold (20\%) was used for testing. Three cases were considered for training, \textbf{Case IV:} Train on 80\% of AV-GPS-Dataset 4, and test on 20\% of AV-GPS-Dataset 4, AV-GPS-Dataset 5, and AV-GPS-Dataset 6; \textbf{Case V:} Train on 80\% of AV-GPS-Dataset 5, and test on AV-GPS-Dataset 4, 20\% of AV-GPS-Dataset 5 and AV-GPS-Dataset 6; \textbf{Case VI:} Train on 80\% of AV-GPS-Dataset 6, and test on AV-GPS-Dataset 4, AV-GPS-Dataset 5, and 20\% of AV-GPS-Dataset 6. For all three cases, the performance of the machine learning models is measured in terms of Accuracy, Precision, Recall, and F1-score. For each model, we used the same tuned hyperparameters utilized as Experiment 3. The performance of machine learning algorithms on different simulation generated datasets is presented in Table \ref{tab2_SimRes}. The results indicate that the performance varies across different datasets as the training set changes. MLP consistently performs with F1 scores above 93\% in all three cases, achieving the highest average F1 score of 97.1\%. Comparing Table \ref{tab2} and Table \ref{tab2_SimRes}, we can see that the GPS-IDS approach demonstrates an overall better performance on the simulation-based datasets.

%%%%%%%%%%%%%_BEGIN_TABLE_%%%%%%%%%%%%%%%%

\begin{table*}
\caption{Performance of the Machine Learning Classification Models on AV-GPS-Datasets 4---6 (Simulation-generated Data)}\label{tab2_SimRes}
  \centering
  \begin{tabular}{|c|c|c|c|c|c|c|c|c|} 
    \hline
    & \textbf{AV-GPS-Datasets 4---6} & \textbf{Metrics} & \textbf{RF} & \textbf{XGB} & \textbf{SVC} & \textbf{MLP} & \textbf{Adaboost} & \textbf{GB}  \\
    \hline
    \multirow{10}{*}{\STAB{\rotatebox[origin=c]{90}{\textbf{Case IV}}}}
    \multirow{10}{*}{\STAB{\rotatebox[origin=c]{90}{Trained with 80\% of Dataset 4}}}
    & \multirow{4}{*}{Test on 20\% of Dataset 4} 
    
    & Accuracy &  0.8724 & 0.9495 & 0.9091& 0.9471 & 0.9283 & 0.9285    \\ 

    & & Precision & 0.8955 & 0.9541 & 0.9093 & 0.9507 & 0.9294 & 0.9295    \\
    
    & & Recall & 0.8724 & 0.9495 & 0.9091 & 0.9471 & 0.9283 & 0.9285  \\ 
    
    & & F1 Score & 0.8705  & 0.9494 & 0.9091 & 0.9470 & 0.9283 & 0.9284   \\ 
    
    \cline{2-9}

    & \multirow{4}{*}{Test on Dataset 5} & Accuracy & 0.9494 & 0.9494  & 0.9523 & 0.9790 & 0.9486 & 0.9068   
    \\ 
    
    & & Precision & 0.9541 & 0.9541 & 0.9564 & 0.9798 & 0.9534 & 0.9215    \\ 
    
    & & Recall & 0.9494 & 0.9494 & 0.9523 & 0.9790 & 0.9486 & 0.9068   \\ 
    
    & & F1 Score & 0.9493  & 0.9493 & 0.9522 & 0.9789 & 0.9485 & 0.9060 \\ 
    \cline{2-9}

    & \multirow{4}{*}{Test on Dataset 6} & Accuracy & 0.9492 & 0.9492 & 0.9527 & 0.9518 & 0.9500 & 0.9151  \\ 

    & & Precision & 0.9539 & 0.9539 & 0.9568 & 0.9560 & 0.9545 & 0.9275    \\ 
    
    & & Recall & 0.9492 & 0.9492 & 0.9527 & 0.9518 & 0.9500 & 0.9151   \\ 
    
    & & F1 Score & 0.9491 & 0.9491 & 0.9526 & 0.9517 & 0.9499 & 0.9145     \\

    \hline

    \multicolumn{3}{|c|}{Average F1 Score in Case I} & 0.9230   &  0.9493 & 0.9380 & \textbf{0.9592} & 0.9422 & 0.9163   \\

    \hline

    %%%%%%%%%%%%%%%%%%%%%%%%%%%%%%%%%%%%%%%%%%%%%%%%%%%%%%%%%%%
    \multirow{10}{*}{\STAB{\rotatebox[origin=c]{90}{\textbf{Case V}}}}
    \multirow{10}{*}{\STAB{\rotatebox[origin=c]{90}{Trained with 80\% of Dataset 5}}}
    & \multirow{4}{*}{Test on Dataset 4} & Accuracy & 0.9409 & 0.9454 & 0.9590 & 0.9619 & 0.9040 & 0.9835   \\
    
    & & Precision & 0.9471 & 0.9493 & 0.9621 & 0.9646 & 0.9040 & 0.9837   \\
    
    & & Recall & 0.9409 & 0.9454 & 0.9590 & 0.9619 & 0.9040 & 0.9835   \\
    
    & & F1 Score & 0.9407 & 0.9453 & 0.9589 & 0.9618 & 0.9040 & 0.9835   \\
    \cline{2-9}

    & \multirow{4}{*}{Test on 20\% of Dataset 5} & Accuracy & 0.8591 & 0.9397 & 0.9492 & 0.9379 & 0.8911 & 0.9285     \\ 

    & & Precision & 0.8760 & 0.9427 & 0.9538 & 0.9408 & 0.8981 & 0.9295  \\
    
    & & Recall & 0.8591 & 0.9397 & 0.9492 & 0.9379 & 0.8911 & 0.9285  \\ 
    
    & & F1 Score & 0.8575 & 0.9396 & 0.9491 & 0.9378 & 0.8906 & 0.9284   \\ 
    \cline{2-9}

    & \multirow{4}{*}{Test on Dataset 6} & Accuracy & 0.9461 & 0.9492 & 0.9987 & 0.9996 & 0.9635 & 0.9151   \\ 
    
    & & Precision & 0.9513 & 0.9539 & 0.9987 & 0.9996 & 0.9660 &   0.9275  \\ 
    
    & & Recall & 0.9461 & 0.9492 & 0.9987 & 0.9996 & 0.9635 & 0.9151    \\ 
    
    & & F1 Score & 0.9459 & 0.9491 & 0.9987 & 0.9996 & 0.9635 & 0.9145    \\
    \hline

    \multicolumn{3}{|c|}{Average F1 Score in Case II} & 0.9147 & 0.9450 & \textbf{0.9689} & 0.9664 & 0.9194 & 0.9421   \\
    
    \hline

    %%%%%%%%%%%%%%%%%%%%%%%%%%%%%%%%%%%%%%%%%%%%%%%%%
    
    \multirow{10}{*}{\STAB{\rotatebox[origin=c]{90}{\textbf{Case VI}}}}
    \multirow{10}{*}{\STAB{\rotatebox[origin=c]{90}{Trained with 80\% of Dataset 6}}}
    & \multirow{4}{*}{Test on Dataset 4} & Accuracy & 0.9429 & 0.9400 & 0.9599 &    0.9721   & 0.9409 & 0.9511   \\ 
    
    & & Precision & 0.9488 & 0.9430 & 0.9628 &  0.9733    & 0.9432 & 0.9554   \\ 
    
    & & Recall & 0.9429 & 0.9400 & 0.9599 &   0.9721   & 0.9409 & 0.9511  \\ 
    
    & & F1 Score & 0.9427 & 0.9399 & 0.9598 &   0.9721   & 0.9409 & 0.9511    \\ 
    \cline{2-9}

    & \multirow{4}{*}{Test on Dataset 5} & Accuracy & 0.9986 & 0.9494 & 0.9998 &   0.9997    & 0.9995 & 0.9962   \\ 
    
    & & Precision & 0.9986 & 0.9541 & 0.9998 &  0.9997  & 0.9995 & 0.9963  \\ 
    
    & & Recall & 0.9986 & 0.9494 & 0.9998 &   0.9997  & 0.9995 & 0.9962   \\ 
    
    & & F1 Score & 0.9986 & 0.9493 & 0.9998 &   0.9997   & 0.9995 & 0.9962   \\ 
    \cline{2-9}

    & \multirow{4}{*}{Test on 20\% of Dataset 6} & Accuracy & 0.8268 & 0.9090 & 0.9522 &   0.9910  & 0.9349 & 0.9107     \\ 
    
    & & Precision & 0.8602 & 0.9229 & 0.9564 &  0.9912    & 0.9360 & 0.9214   \\ 
    
    & & Recall & 0.8268 & 0.9090 & 0.9522 &   0.9910    & 0.9349 & 0.9107     \\ 
    
    & & F1 Score & 0.8227 & 0.9082 & 0.9521 &   0.9910   & 0.9348 & 0.9101    \\
    \hline

    \multicolumn{3}{|c|}{Average F1 Score in Case III} & 0.9213 & 0.9325 & 0.9706 &  \textbf{0.9876}  & 0.9584 & 0.9525     \\

    \hline

    \multicolumn{3}{|c|}{\textbf{Average F1 Score of all three cases}} & 0.9197 & 0.9423 & 0.9592 &  \textbf{0.9711}  & 0.9400 & 0.9370   \\
    
    \hline

  \end{tabular}
\end{table*}

\subsubsection{Experiment 7- Tuning of the Detection Threshold $\mathbb{T}$ on Simulated Data}

\begin{figure}[t]
\centerline{\includegraphics[width=7.5cm]{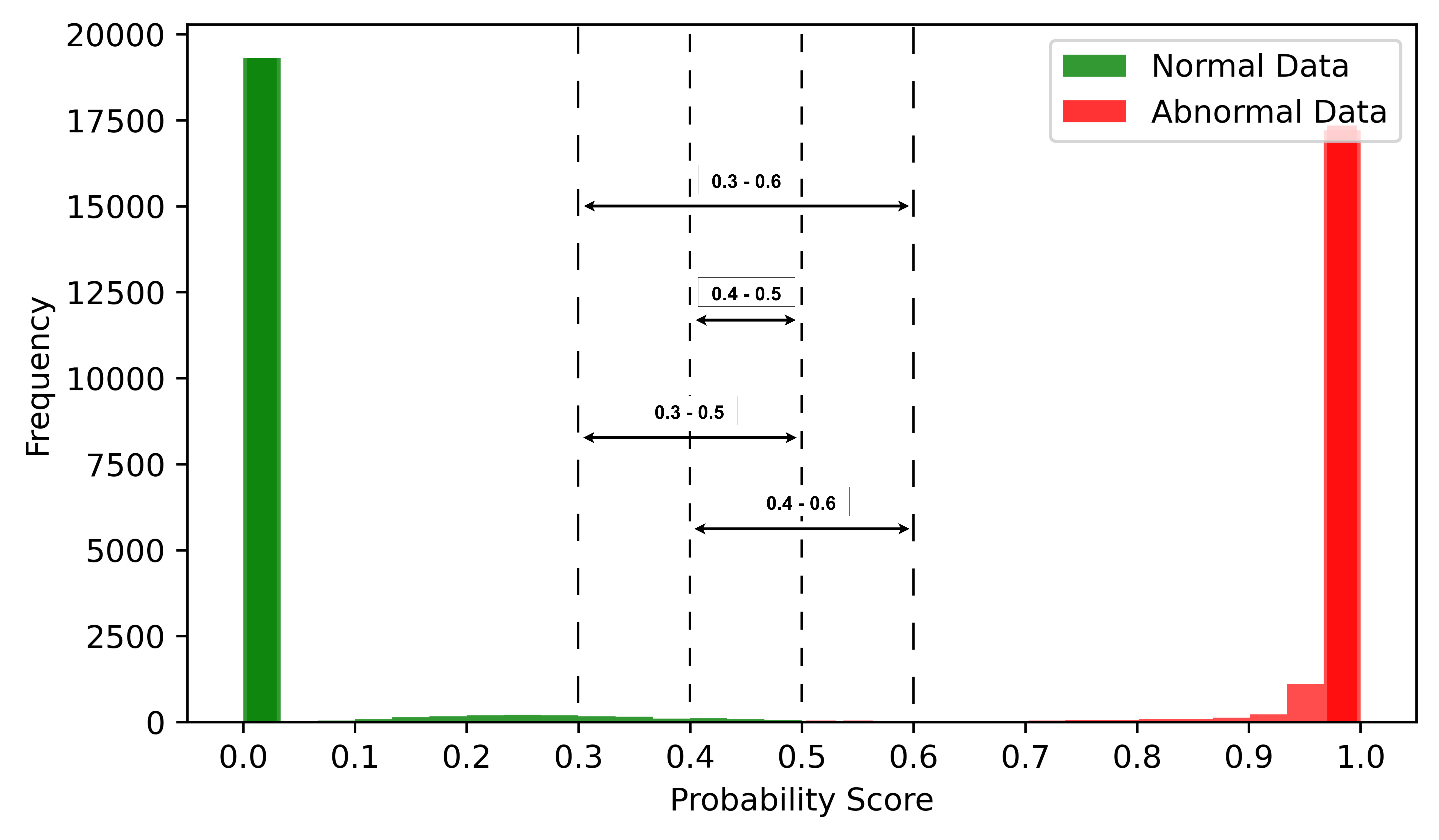}}
\caption{Comparison of probability score distribution for normal and attack data in simulations}\label{prob_score_margin_sim}
\end{figure}

Similar to Experiment 4, this experiment assessed the sensitivity of the GPS-IDS detection threshold $\mathbb{T}$. We varied threshold values to evaluate their impacts on GPS-IDS performance using the simulation-generated datasets. The machine learning model used for this experiment was the MLP, given its consistent best performance in the previous experiments. The resulting data were plotted to determine a suitable detection margin, as illustrated in Fig. \ref{prob_score_margin_sim} %Fig. \ref{prob_score_margin_sim} depicts the probability score distributions for normal and attack data obtained by applying Case I and Case II training methodologies on the same graph. 
By comparing the probability score distributions for normal and attack data, %it becomes evident that the two kinds of data can be effectively distinguished with a margin indicated by the dotted bars. 
we again find that selecting a detection threshold $\mathbb{T}$ within a range of 0.4 to 0.5 shows the best performance, resulting in misclassification of only 295 instances of normal data and no instance of attack data and leading to a False Positive Rate of 0.0076 or 0.76\% and False Negative Rate of 0\%. Table. \ref{FPFN_sim} summarizes the different False Positive and False Negative rates for different margins in simulated data.

\begin{table}[t]
\caption{False Positive (FP) and False Negative (FN) Rates for Different Detection Margins in Simulated Datasets} \label{FPFN_sim}
\centering
\begin{tabular}{c|c|c|c|c}
\hline
\hline
\multirow{2}{*}{\begin{tabular}[c]{@{}c@{}}Detection\\ Margin \end{tabular}} & \multirow{2}{*}{\begin{tabular}[c]{@{}c@{}}Normal Data\\ Misclassified\end{tabular}} & \multirow{2}{*}{\begin{tabular}[c]{@{}c@{}}Attack Data\\ Misclassified\end{tabular}} & \multirow{2}{*}{\begin{tabular}[c]{@{}c@{}}FP \\ Rate\end{tabular}} & \multirow{2}{*}{\begin{tabular}[c]{@{}c@{}}FN\\ Rate\end{tabular}} \\
 &  &  &  &  \\
%Detection Margin Range & Normal Data Misclassified & Attack Data Misclassified & False Positive Rate & False Negative Rate \\
\hline
\textbf{0.4 - 0.5} & \textbf{295} & \textbf{0} & \textbf{0.0076} & \textbf{0.0} \\
0.3 - 0.5 & 764 & 0 & 0.0198 & 0.0 \\
0.4 - 0.6 & 295 & 200 & 0.0076 & 0.0052 \\
0.3 - 0.6 & 764 & 200 & 0.0198 & 0.0052  \\
\hline
\hline
\end{tabular}
\end{table}

\section{Conclusion and Future Work}\label{conclusion}
This paper presents a novel anomaly behavior analysis-based GPS Intrusion Detection framework called the GPS-IDS that detects abnormal GPS navigation in AVs. The approach uses physics-based vehicle modeling to represent the behavior of the vehicle. A modified dynamic bicycle model is utilized to capture the normal behavior of the vehicle and machine learning techniques are used to detect the anomalous behavior. Moreover, a novel dataset family called the AV-GPS-Datasets %with over 69,000 instances of real-world autonomous vehicle normal data and GPS spoofing attack data
with both real-world and simulated data is introduced in this paper. The performance of the proposed GPS-IDS approach is evaluated using the AV-GPS-Dataset, and the experimental results affirm its high detection rate of GPS spoofing attacks. Additionally, it is validated that the proposed approach exhibits faster detection times in comparison to the EKF-based detection algorithm in real data of AV-GPS-Datasets. In contrast to the existing segregated intrusion detection techniques that concentrate only on individual sensor data, our approach considers the overall behavior of the system, providing a more comprehensive approach to detect GPS attacks. We argue that using GPS-IDS in conjunction with sensor-specific intrusion detection systems can provide a powerful defense mechanism against GPS spoofing attacks in autonomous vehicles. To ensure the generalization of our approach, we conducted simulation-based experiments to show that our suggested system can perform equally well in a more complex urban scenario with multiple vehicles under attack.

Moving forward, there are several promising research avenues to extend this work. The GPS-IDS algorithm can be refined by incorporating additional operational parameters in the Autonomous Vehicle Behavior Model, such as integrating accurate modeling of multiple sensors and more robust control algorithms. Further research can be done to explore the effectiveness of integrating GPS-IDS with existing sensor-specific detection systems to enhance the overall security of AVs. Additionally, GPS-IDS can be improved to detect more stealthy GPS attacks, such as dual-band spoofing and spoofing when aligned with road shapes \cite{ZengAll, NarainSecurity} against AVs. Furthermore, the feasibility of real-world deployment scenarios, including cloud environment computing, onboard deployment on resource-constrained edge devices, or hybrid computation strategies that balance cloud and onboard processing can be investigated to validate the scalability and practicality of the GPS-IDS approach. Finally, the development of a digital twin for the Autonomous Vehicle Behavior Model emerges as a promising direction, providing a virtual environment for comprehensive testing and simulation. 

\vspace{-10pt}

\begin{IEEEbiography}[{\includegraphics[width=1in, height=1.25in, clip, keepaspectratio]{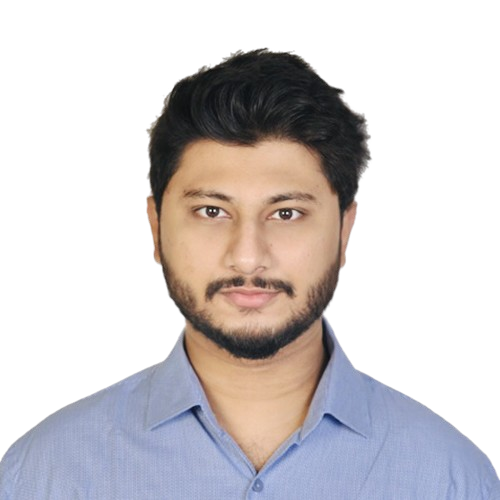}}]{Murad Mehrab Abrar} is currently a Master's student in the Department of Electrical and Computer Engineering, the University of Arizona, Tucson, AZ, USA. He completed his B.Sc. in Electrical and Electronic Engineering from Ahsanullah University of Science and Technology, Dhaka, Bangladesh in 2019. His research interest focuses on robotics, autonomous vehicles, and machine learning.
\end{IEEEbiography}

\vspace{-4pt}

\begin{IEEEbiography} [{\includegraphics[width=1in, height=1.25in, clip, keepaspectratio]{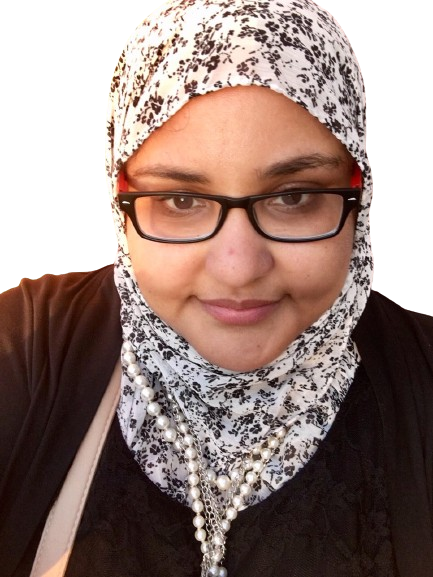}}]{Amal Youssef} is currently a PhD student in the department of Electrical and Computer Engineering, Tucson, AZ, USA. She received her M.Sc. degree from the University of Utah, USA in 2018. Her current research interests include cybersecurity, autonomous vehicles, and machine learning.
    
\end{IEEEbiography}

\vspace{-4pt}

\begin{IEEEbiography}[{\includegraphics[width=1in, height=1.25in, clip, keepaspectratio]{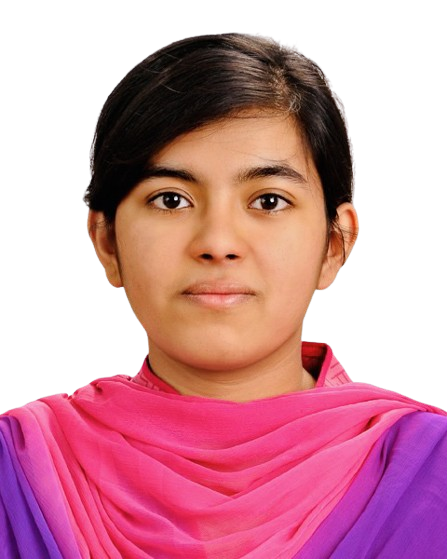}}]{Raian Islam} is a Master’s student in the Department of Electrical and Computer Engineering at the University of Arizona, Tucson, AZ, USA. She received her B.Sc. degree in Electrical and Electronic Engineering from Ahsanullah University of Science and Technology, Dhaka, Bangladesh, in 2019. Her current research interests include data analytics, photovoltaics, and machine learning.

\end{IEEEbiography}

\vspace{-4pt}

\begin{IEEEbiography}[{\includegraphics[width=1in, height=1.25in, clip, keepaspectratio]{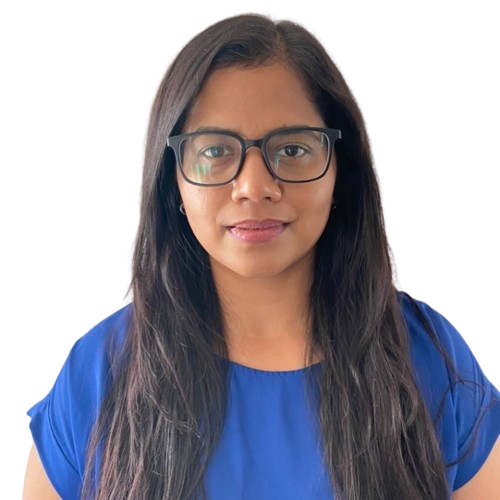}}]{Shalaka Satam} received her B.E. degree in Electronics and Telecommunications Engineering from the University of Mumbai in 2015. She received the M.S. and Ph.D. degrees in Electrical and Computer Engineering from the University of Arizona in 2017 and 2022, respectively. Her research focuses on cybersecurity, autonomous vehicles, and Internet of Things.

\end{IEEEbiography}

\vspace{-4pt}

\begin{IEEEbiography}[{\includegraphics[width=1.2in, height=1.3in, clip, keepaspectratio]{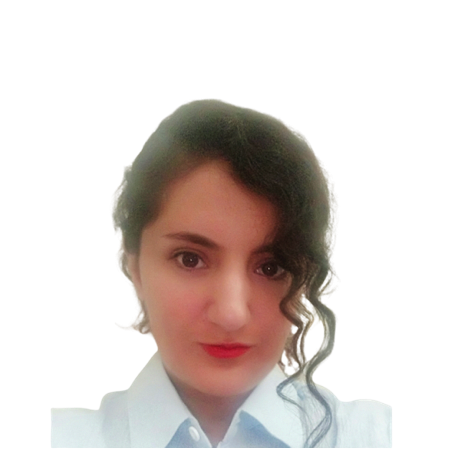}}]{Banafsheh Saber Latibari} received the B.Sc. degree in Computer Engineering from the K. N. Toosi University of Technology and the M.Sc. degree in Computer Architecture from the Sharif University of Technology. She received her Ph.D. degree from the Electrical and Computer Engineering (ECE) Department at the University of California, Davis. She is currently a Postdoctoral Research Associate at the University of Arizona's ECE Department. Her research focuses on deep learning, embedded system security, and computer architecture.
    
\end{IEEEbiography}

\vspace{-4pt}

\begin{IEEEbiography}[{\includegraphics[width=1in, height=1.25in, clip, keepaspectratio]{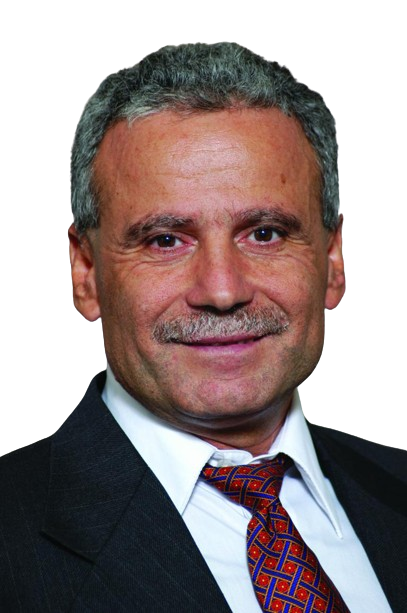}}]{Salim Hariri} (Senior Member, IEEE) received an M.Sc. degree from Ohio State University in 1982, and a Ph.D. degree in Computer Engineering from the University of Southern California in 1986. He is a Professor in the Department of Electrical and Computer Engineering, the University of Arizona, and the Director of the NSF Center for Cloud and Autonomic Computing (NSF-CAC). His research focuses on autonomic computing, cybersecurity, cyber resilience, and cloud security.

\end{IEEEbiography}

\vspace{-4pt}

\begin{IEEEbiography}[{\includegraphics[width=1in, height=1.25in, clip, keepaspectratio]{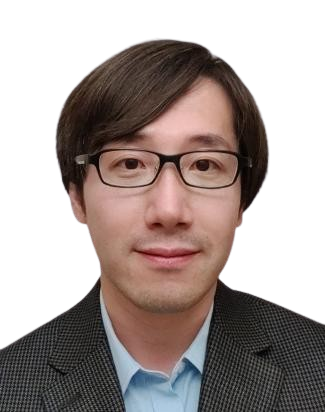}}]{Sicong Shao} is an assistant professor of the School of Electrical Engineering and Computer Science at the University of North Dakota (UND). Before joining UND, he was a research assistant professor in the Department of Electrical and Computer Engineering (ECE) at the University of Arizona where he also received his Ph.D. in ECE.  His research interests include cybersecurity, machine learning, artificial intelligence, and software engineering.

\end{IEEEbiography}

\vspace{-4pt}

\begin{IEEEbiography}[{\includegraphics[width=1in, height=1.25in, clip, keepaspectratio]{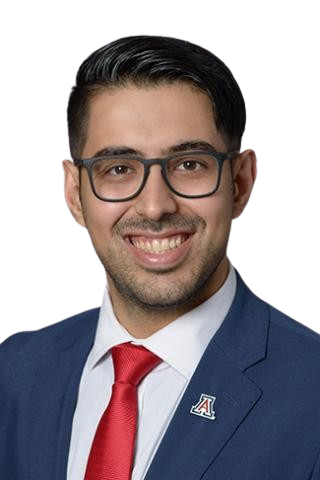}}]{Soheil Salehi} (Member, IEEE) is an assistant professor in the Electrical and Computer Engineering (ECE) Department at the University of Arizona (UofA). Prior to joining the UofA, Soheil was an NSF-Sponsored Computing Innovation Fellow in the Accelerated, Secure, and Energy-Efficient Computing Laboratory and the Center for Hardware and Embedded Systems Security and Trust at the University of California, Davis (UC Davis). He received his Ph.D. and M.S. degrees in ECE from the University of Central Florida (UCF) in 2016 and 2020, respectively. He has expertise in the areas of hardware security and IoT supply-chain security as well as applied ML for secure hardware design. Moreover, he has designed novel circuits and architectures for secure and accelerated computing. He has received several nominations and award recognition, which include the Outstanding Reviewer Award at IEEE/ACM DAC'23, the Best Poster Award at ACM GLSVLSI'19, the Best Paper Award Nominee at IEEE ISQED'17 as well as the Best Presentation at UC Davis Postdoctoral Research Symposium in 2021, and the Best Graduate Teaching Assistant Award at UCF in 2016. 
\end{IEEEbiography}

\vspace{-4pt}

\begin{IEEEbiography}[{\includegraphics[width=1in, height=1.25in, clip, keepaspectratio]{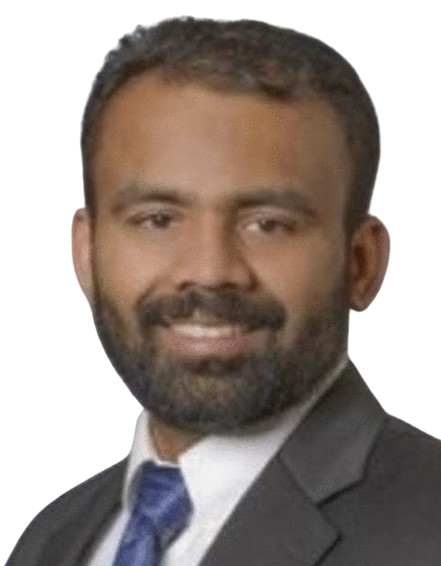}}]{Pratik Satam} is an Assistant Professor in the Department of Systems and Industrial Engineering at the University of Arizona, Tucson, AZ, USA. He is also part of the new Software Engineering program in College of Engineering at the University of Arizona. He received his B.E. degree in Electronics and Telecommunication Engineering from the University of Mumbai, in 2013, and the M.S. and Ph.D. degrees in Electrical and Computer Engineering from the University of Arizona in 2015 and 2019, respectively. From 2019 to 2022, he has been a Research Assistant Professor at the Department of Electrical and Computer Engineering, the University of Arizona. His current research interests include autonomic computing, cyber security, cyber resilience, secure critical infrastructures, and cloud security. He is an Associate Editor for the scientific journal Cluster Computing.

\end{IEEEbiography}

%Use $\backslash${\tt{begin\{IEEEbiography\}}} and then for the 1st argument use $\backslash${\tt{includegraphics}} to declare and link the author photo. Use the author name as the 3rd argument followed by the biography text.

%\end{IEEEbiography}



\vfill

\end{document}